\documentclass[11pt, a4paper]{article}

\usepackage[margin=32mm]{geometry}

\usepackage{amssymb,amsmath,amsfonts,amsthm, amscd, mathrsfs,helvet, mathtools}
\usepackage{bm, stmaryrd}
\usepackage{graphicx,verbatim}
\usepackage[usenames, dvipsnames]{color}
\usepackage[mathcal]{euscript}
\usepackage{psfrag}
\usepackage[all]{xy}
\usepackage{tikz}
\usetikzlibrary{arrows,matrix,decorations.markings,shapes.geometric}
\usepackage[normalem]{ulem}

\theoremstyle{plain}
\newtheorem{theorem}{Theorem}
\newtheorem*{theorem*}{Theorem}
\newtheorem{proposition}[theorem]{Proposition}
\newtheorem*{proposition*}{Proposition}

\theoremstyle{definition}
\newtheorem*{example}{Example}

\newcommand{\tensor}[1]{{\mathfrak{#1}}}

\newcommand{\longhookrightarrow}{\lhook\joinrel\relbar\joinrel\rightarrow}
\newcommand{\longtwoheadrightarrow}{\relbar\joinrel\twoheadrightarrow}

\newcommand{\gray}[0]{\color{Gray}}
\definecolor{brightRed}{rgb}{1,0,0}
\newcommand{\red}[0]{\color{brightRed}}
\definecolor{brightBlue}{rgb}{0,0,1}
\newcommand{\blue}[0]{\color{brightBlue}}
\definecolor{myGreen}{rgb}{0,0.7,0}
\newcommand{\green}[0]{\color{myGreen}}

\DeclareMathOperator{\res}{res}

\def\Loop{\mathcal{L}}
\def\a{\mathfrak{a}}
\def\b{\mathfrak{b}}

\def\d{\mathfrak{d}}

\def\g{\mathfrak{g}}

\def\h{\mathfrak{h}}
\def\k{\mathfrak{k}}

\def\n{\mathfrak{n}}
\def\p{\mathfrak{p}}

\def\F{f}

\def\ha{\mbox{\small $\frac{1}{2}$}}
\def\qa{\mbox{\small $\frac{1}{4}$}}

\def\CC{\mathbb{C}}
\def\RR{\mathbb{R}}

\def\J{\mathcal{J}}
\def\Jkm{J}

\def\R{\mathscr{R}}
\def\L{\mathscr{L}}

\def\1{\tensor{1}}
\def\2{\tensor{2}}
\def\3{\tensor{3}}
\def\4{\tensor{4}}

\numberwithin{equation}{section}

\begin{document}

\begin{center}
\vspace*{2em}
{\Large\bf Deformed integrable $\sigma$-models, classical $R$-matrices}\\[2mm]
{\Large \bf and classical exchange algebra on Drinfel'd doubles}\\
\vspace{1.5em}
{\large Beno\^{\i}t Vicedo}

\vspace{1em}
\begingroup\itshape
{\it School of Physics, Astronomy and Mathematics,
University of Hertfordshire,}\\
{\it College Lane,
Hatfield AL10 9AB,
United Kingdom}
\par\endgroup
\vspace{1em}
\begingroup\ttfamily
Benoit.Vicedo@gmail.com
\par\endgroup
\vspace{1.5em}
\end{center}

\begin{abstract}
We describe a unifying framework for the systematic construction of integrable deformations of integrable $\sigma$-models within the Hamiltonian formalism. It applies equally to both the `Yang-Baxter' type as well as `gauged WZW' type deformations which were considered recently in the literature. As a byproduct, these two families of integrable deformations are shown to be Poisson-Lie $T$-dual of one another.
\end{abstract}

\section{Introduction}

Given an integrable $\sigma$-model, it is interesting to ask which kind of deformations, if any, preserve its integrability. Such integrable deformations have recently been the subject of intensive study, driven most notably by their relevance in the context of the AdS/CFT correspondence.
In particular, two families of integrable deformations of a very different nature have been actively developed over the past two years, both of which have been successfully applied to the $AdS_5 \times S^5$ superstring.

\medskip

The first family of integrable deformations was originally introduced by {Klim$\check{\text{c}}$\'{\i}k} in the case of the principal chiral model on an arbitrary (compact) real Lie group $G$ \cite{Klimcik:2002zj, Klimcik:2008eq}. This type of deformation is constructed with the help of a solution $R \in \text{End}\, \g$ of the modified classical Yang-Baxter equation (mCYBE) on $\g = \text{Lie}(G)$, which reads
\begin{equation} \label{mCYBE intro}
[RX, RY] - R \big( [RX, Y] + [X, RY] \big) = - c^2 [X, Y],
\end{equation}
for any $X, Y \in \g$, and where $c^2 \in \mathbb{R}$.
More precisely, solutions of the mCYBE on a real Lie algebra $\g$ fall into three distinct classes, depending on whether
\begin{equation*}
c^2 < 0, \qquad\quad
c^2 > 0 \qquad \text{or} \qquad
c = 0.
\end{equation*}
The deformation of the principal chiral model, called the Yang-Baxter $\sigma$-model, whose integrability was proved in \cite{Klimcik:2008eq} corresponds to the case $c^2 < 0$, \emph{i.e.} $c \in i \mathbb{R} \setminus \{ 0 \}$.
Recently, this type of deformation with $c^2 < 0$ was generalised to all symmetric space $\sigma$-models \cite{Delduc:2013fga} as well as to semi-symmetric space $\sigma$-models \cite{Delduc:2013qra, Delduc:2014kha} on the example of the $AdS_5 \times S^5$ superstring. The deformation parameter was originally called $\eta$ (and later $\varkappa$ in \cite{Hoare:2014pna}) and so such deformations have come to be known as $\eta$-deformations or $\varkappa$-deformations in the literature. Sometimes these are also referred to as $q$-deformations since the global symmetry algebra of these theories was shown to be $q$-deformed \cite{Delduc:2013fga} with $q$ a function of $\eta$. We prefer to refer to these as `Yang-Baxter' type deformations instead to reflect the fact that they are built from solutions of the (modified) classical Yang-Baxter equation. The target space geometry of the Yang-Baxter type deformation of the $AdS_5 \times S^5$ superstring and its properties have been extensively studied -- see \emph{e.g.} \cite{Arutyunov:2013ega, Hoare:2014pna, Arutynov:2014ota, Lunin:2014tsa, Engelund:2014pla}. Although the case $c^2 > 0$, namely $c \in \mathbb{R} \setminus \{ 0 \}$, has received comparatively little attention, Yang-Baxter type deformations certainly exist for solutions of the mCYBE in all three of the above
cases \cite{Klimcik:2002zj, Matsumoto:2015jja}. In particular, Yang-Baxter deformations with $c = 0$, known as the CYBE case, have also been studied at great length in the context of the AdS/CFT correspondence -- see \emph{e.g.} \cite{Matsumoto:2015jja, Kawaguchi:2014qwa, Matsumoto:2014cja}. A number of important issues regarding this class of deformations were clarified very recently in \cite{vanTongeren:2015soa}. It was shown, in particular, that all know examples of such deformations of  the $AdS_5 \times S^5$ superstring arise from solutions of \eqref{mCYBE intro} with $c=0$ on the real Lie algebra $\mathfrak{su}(2,2|4)$ without the need to complexify.

\medskip

More recently, a second family of integrable deformations was proposed by Sfetsos in the case of the non-abelian $T$-dual of the principal chiral model on an arbitrary Lie group $G$ \cite{Sfetsos:2013wia}, building on earlier work of Balog \emph{et al.} \cite{Balog:1993es} on deformations of the $SU(2)$ principal chiral model. An elegant construction of these deformations is performed at the level of the action by starting from the action of a $G$-valued principal chiral field, adding to it the action for another $G$-valued WZW field, and then gauging the resulting sum. In a certain gauge this action describes the deformation of a $G/G$ gauged WZW model.
Such deformations are commonly referred to as $\lambda$-deformations, due to the name given to the deformation parameter in this case. We prefer here as well to refer to these as `gauged WZW' type deformations to reflect the nature of the deformation. Gauged WZW type deformations were recently generalised to all symmetric space $\sigma$-models \cite{Hollowood:2014rla} and to semi-symmetric space $\sigma$-models in the case of the $AdS_5 \times S^5$ superstring \cite{Hollowood:2014qma}. Very recently, these deformations and in particular their target space geometries were further analysed in \cite{Demulder:2015lva}.

\medskip

A natural question at this point is whether the above two families of deformations are related in some way or another. In order to address this question, it is necessary first to have a description of both deformations within the same framework. The Hamiltonian formalism seems the most appropriate for this. Indeed, the systematic construction of Yang-Baxter type deformations was already achieved in \cite{Delduc:2013fga, Delduc:2014kha} using integrability based methods within the Hamiltonian formalism. One immediate advantage of this constructive approach to integrable deformations is that the models obtained are automatically integrable. In stark contrast to this, the gauged WZW type deformations are presently constructed at the level of the action, so their integrability is not immediately apparent. One of the aims of the present paper is to give a more constructive definition of these deformations within the Hamiltonian formalism. Furthermore, another advantage of using the Hamiltonian formalism is that it is the appropriate setting for determining the algebra of hidden symmetries of the deformed model.

\medskip

In this paper we will therefore focus only on the Hamiltonian aspects of deformations. Specifically, we will reformulate the construction of Yang-Baxter type deformations of \cite{Delduc:2013fga} in a language suitable for describing also gauged WZW type deformations. We shall only be concerned with field theories for which the phase space is given by the cotangent bundle $T^\ast \Loop G$ of the loop group $\Loop G \coloneqq C^\infty(S^1, G)$ of some real Lie group $G$. We assume that the initial `undeformed' theory is integrable with a Lax matrix $\L(\lambda, \theta)$ which is a function of the phase space fields $g \in \Loop G$, $X \in \Loop \g$ parameterising $T^\ast \Loop G$ and depending rationally on the complex spectral parameter $\lambda \in \CC$. Furthermore, we assume that the Poisson bracket of the Lax matrix with itself is of the general non-ultralocal type \cite{Maillet86}. The latter depends on a rational function $\varphi(\lambda)$, called the `twist function', which is the central object in the construction of \cite{Delduc:2013fga}. Indeed, deformations of the given integrable model were constructed in \cite{Delduc:2013fga} simply by deforming the twist function.

\medskip

\begin{figure}[h]
\centering
\def\svgwidth{90mm}
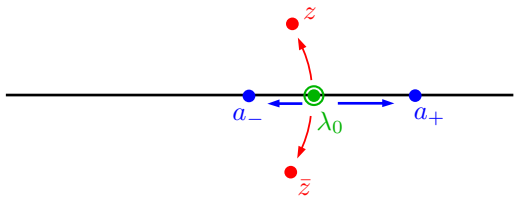
\caption{Behaviour of the poles of twist function in the real (blue), complex (red) and CYBE (green) branches.}
\label{fig: poles coset}
\end{figure}

Starting from an undeformed model whose twist function $\varphi(\lambda)$ has a double pole at some point $\lambda_0 \in \mathbb{R}$, there are two ways of deforming the twist function while preserving its reality conditions. We can deform the double pole at $\lambda_0$ into a pair of simple poles which are either both real or complex conjugate of one another.
These will be referred to as the real and complex branches, respectively. However, deforming the twist function is not enough to define a new $\sigma$-model. The key to defining an actual deformation is to extract from the Lax matrix and the deformed twist function a pair of fields $g \in \Loop G$ and $X \in \Loop \g$ parametrising the cotangent bundle $T^\ast \Loop G$. The extra input needed to achieve this is a solution $R \in \text{End}\, \g$ of the mCYBE on the Lie algebra $\g$ \cite{SemenovTianShansky:1983ik}. Specifically, in the real (resp. complex) branch, we require a solution of \eqref{mCYBE intro} with $c^2 > 0$ (resp. $c^2 < 0$). There is however one further type of deformation possible. If the twist function is not modified, in particular the double pole at $\lambda_0$ is unchanged, we can still use a solution of \eqref{mCYBE intro} with $c = 0$ to construct an interesting deformation. We shall refer to this as the CYBE branch. In particular, we see that there is a one-to-one correspondence between the three classes of solutions of the mCYBE on the real Lie algebra $\g$ and the three ways of `deforming' the double pole $\lambda_0$ of the twist function. See Figure \ref{fig: poles coset} for a schematic representation of the behaviour of the double pole $\lambda_0$ of the twist function in each of the three branches.

The deformations of most interest will be those in the real and complex branches. The corresponding solutions of the mCYBE with $c^2 > 0$ and $c^2 < 0$, respectively, have a simple algebraic interpretation in terms of Drinfel'd doubles. Specifically, in the case $c^2 < 0$ there is a one-to-one correspondence
\begin{align*}
\left\{ \begin{array}{c}
\text{solutions of mCYBE}\\
\text{on $\g$ with $c^2 < 0$}
\end{array} \right\} \quad &\overset{1\--1}\longleftrightarrow \quad
\left\{ \begin{array}{c}
\text{subalgebras $\p \subset \g^{\CC}$}\\
\text{with $\g^\CC = \g \dotplus \p$}
\end{array} \right\},
\end{align*}
where $\g^\CC = \g \otimes_{\mathbb{R}} \CC$ is the complexification of $\g$.
Likewise, for the case $c^2 > 0$ we have the one-to-one correspondence
\begin{align*}
\left\{ \begin{array}{c}
\text{solutions of mCYBE}\\
\text{on $\g$ with $c^2 > 0$}
\end{array} \right\} \quad &\overset{1\--1}\longleftrightarrow \quad
\left\{ \begin{array}{c}
\text{subalgebras $\p \subset \d$}\\
\text{with $\d = \g^{\delta} \dotplus \p$}
\end{array} \right\},
\end{align*}
where $\d = \g \oplus \g$ is the so called real double of $\g$ and $\g^\delta$ is the diagonal subalgebra in $\d$. The Lie algebras $\g^\CC$ and $\d$ are both examples of Drinfel'd doubles $\g \oplus \g^\ast$ of the real Lie algebra $\g$, corresponding to different choices of Lie brackets on the dual $\g^\ast$. The vector space decompositions $\g^\CC = \g \dotplus \p$ or $\d = \g^\delta \dotplus \p$ at the level of the Lie algebras induce factorisations at the Lie group level. Specifically, if we denote by $\mathcal D$ the connected and simply connected Lie group corresponding to either the Lie algebra $\g^\CC$ or $\d$, by $G \subset \mathcal D$ the subgroup with Lie algebra $\g$ or $\g^{\delta}$ and by $G^\ast \subset \mathcal D$ the subgroup with Lie algebra $\p$ in the above notation, then in the simplest of cases we obtain a bicrossproduct factorisation $\mathcal D = G G^\ast = G^\ast G$.

\medskip

We now describe how a choice of solution of the mCYBE with $c^2 > 0$ (resp. $c^2 < 0$) enables us to extract fields $g \in \Loop G$ and $X \in \Loop \g$ parametrising the (left) trivialisation of $T^\ast \Loop G$ in the real (resp. complex) branch. Consider the value of the Lax matrix at the pair of simple poles of the twist function. In the complex branch we have $\L(z, \theta) \in \g^\CC$ since $z \in \CC$ (and $\L(\bar z, \theta)$ is related to it by conjugation), whereas in the real branch we have $\big( \L(a_+, \theta), \L(a_-, \theta) \big) \in \d$ since $a_+$ and $a_-$ are both real. Introducing the extended solution $\Psi(\lambda, \theta)$ through the relation
\begin{equation} \label{ext sol intro}
- \partial_\theta \Psi(\lambda, \theta) \Psi(\lambda, \theta)^{-1} = \L(\lambda, \theta),
\end{equation}
it follows that $\Psi(z, \theta) \in G^\CC$ in the complex branch and $\big( \Psi(a_+, \theta), \Psi(a_-, \theta) \big) \in D$ in the real branch. By factorising these fields valued in the Drinfel'd double $\mathcal D$, we obtain the sought after pair of fields $g \in \Loop G$ and $X \in \Loop \g$
parameterising the (left) trivialisation $\Loop G \times \Loop \g$ of the cotangent bundle $T^\ast \Loop G$. More precisely, in the real branch, depending on which order we choose the factors in, namely $\mathcal D = G G^\ast$ or $\mathcal D = G^\ast G$, we obtain either the fields of the Yang-Baxter type deformation or those of the gauged WZW type deformation. This unifying description of both types of deformations shows that they are Poisson-Lie $T$-dual \cite{Klimcik:1995ux, Klimcik:1995dy}. In the complex branch, the factorisation $\mathcal D = G G^\ast$ leads to the usual Yang-Baxter type deformation but the reverse factorisation $\mathcal D = G^\ast G$ does not lead naturally to a gauged WZW type model.

\medskip

So far we have only described how to extract the pair of fields parameterising $T^\ast \Loop G$ from the Lax matrix. In order to show that their Poisson brackets are the canonical ones on $T^\ast \Loop G$ we use the following general result which is proved in Proposition \ref{prop: KM currents}: at any pair of simple poles $z \neq z'$ of the twist function $\varphi(\lambda)$, the values $\L(z, \theta)$ and $\L(z', \theta)$ of the Lax matrix form a pair of Poisson commuting Kac-Moody currents. Note that the appearance of such Kac-Moody currents was first observed by Rajeev \cite{Rajeev:1988hq} in the context of a single-parameter deformation of the principal chiral model. A similar observation was made in \cite{Hollowood:2014rla} for a one-parameter deformation of symmetric space $\sigma$-models and in the case of a two-parameter deformation of the principal chiral model in \cite{Delduc:2014uaa}. Therefore Proposition \ref{prop: KM currents} can  be seen as a generalisation of these results to any integrable $\sigma$-model with Poisson brackets of the non-ultralocal type \cite{Maillet86} whose the twist function has simple poles. It follows from this that the values $\Psi(z, \theta)$ and $\Psi(z', \theta)$ of the extended solution at the simple poles $z \neq z'$ satisfy a classical `exchange algebra' of the form
\begin{align*}
\{ \Psi_{\1}(z, \theta), \Psi_{\2}(z, \theta') \} &= \gamma \Psi_{\1}(z, \theta) \Psi_{\2}(z, \theta') \big( R_{\1\2} + c\, C_{\1\2} \epsilon_{\theta \theta'} \big),\\
\{ \Psi_{\1}(z, \theta), \Psi_{\2}(z', \theta') \} &= 0,\\
\{ \Psi_{\1}(z', \theta), \Psi_{\2}(z', \theta') \} &= \gamma' \Psi_{\1}(z', \theta) \Psi_{\2}(z', \theta') \big( R_{\1\2} + \bar c \, C_{\1\2} \epsilon_{\theta \theta'} \big),
\end{align*}
where $R_{\1\2}$ is the kernel of the solution $R \in \text{End}\, \g$ of the mCYBE introduced above and $\gamma$, $\gamma'$ are related to the deformation parameter(s). After factorising $\Psi(z, \theta)$ and $\Psi(z', \theta)$ in the Drinfel'd double, these classical exchange relations descend to the appropriate Poisson brackets on the cotangent bundle $T^\ast \Loop G \simeq \Loop G \times \Loop \g$. When $\gamma' = - \gamma$ and $c^2 > 0$ so that $c$ is real, the above Poisson brackets are exactly those satisfied by a pair of left and right chiral WZW fields \cite{BDF90,AS90,F90,G91, Falceto:1992bf}. In the complex branch the interpretation of the above Poisson bracket as a chiral WZW phase space is less clear.

\medskip

Finally, let us comment briefly on deformations in the CYBE branch. In this branch the deformed model is defined by the requirement that after a suitable canonical transformation by a non-local field we recover the original undeformed model. This canonical transformation is defined with the help of a solution to \eqref{mCYBE intro} with $c = 0$. What makes such a deformation non-trivial is that although it can be undone by a canonical transformation, the latter generically introduces a non-trivial twist in the boundary conditions of the $G$-valued field $g$. Such a phenomenon was first observed in the context of strings propagating on the TsT-transformed $AdS_5\times S^5$ background \cite{Frolov:2005dj, Alday:2005ww}, which was later shown to belong to the class of Yang-Baxter type deformations \cite{Matsumoto:2014nra}.

\medskip

The paper is organised as follows. In section \ref{sec: R-matrix} we begin with a review of some well known facts about solutions of the mCYBE on real Lie algebras. For the convenience of the reader we include some of the proofs. Section \ref{sec: def} describes the setup for constructing integrable deformations of field theories on the cotangent bundle $T^\ast \Loop G$. In particular, we discuss the undeformed model in the spirit of the rest of the paper, showing how to construct it from the Lax matrix and twist function alone. In sections \ref{sec: type I}, \ref{sec: type II} and \ref{sec: type III} we discuss the construction of integrable deformations in the complex, real and CYBE branches respectively. In particular, in the real branch case we construct both the Yang-Baxter type and gauged WZW type deformations. The construction is illustrated on the example of the symmetric space $\sigma$-model but all the relevant results are proved in generality. We also exemplify the construction on a two parameter deformation of the principal chiral model. We end with some concluding remarks.

\section{$R$-matrices on real Lie algebras} \label{sec: R-matrix}

Let $\g$ be a real Lie algebra equipped with a non-degenerate invariant symmetric bilinear form $\langle \cdot, \cdot \rangle : \g \times \g \to \mathbb{R}$. Let $G$ be the connected and simply connected real Lie group corresponding to the Lie algebra $\g$.

To any $\mathbb{R}$-linear operator $R \in \text{End}\, \g$ we associate a bilinear, skew-symmetric operation $[\cdot, \cdot]_R : \g \times \g \to \g$, called the $R$-bracket, defined by
\begin{equation} \label{R-bracket}
[X, Y]_R \coloneqq [RX, Y] + [X, RY],
\end{equation}   
for any $X, Y \in \g$. The linear operator $R$ is referred to as an $R$-matrix if \eqref{R-bracket} defines a Lie bracket on $\g$, in other words if it satisfies the Jacobi identity. In this case it is conventional to denote by $\g_R$ the vector space $\g$ equipped with the Lie bracket \eqref{R-bracket}. A sufficient condition for \eqref{R-bracket} to satisfy the Jacobi identity is given by the modified classical Yang-Baxter equation \cite{SemenovTianShansky:1983ik, Semenov-Tian-Shansky2}
\begin{equation} \label{mCYBEc}
[RX, RY] - R\big( [X, Y]_R \big) = - c^2 [X, Y],
\end{equation}
where $c^2 \in \mathbb{R}$. To emphasise the dependence on the parameter $c$, we will refer to this equation as mCYBE$(c)$. Note that it can be rewritten in the following more suggestive form
\begin{equation} \label{mCYBE}
(R \pm c) \big( [X, Y]_R \big) = \big[ (R \pm c) X, (R \pm c) Y \big].
\end{equation}
If $c \neq 0$ then by rescaling the $R$-matrix by $1/|c|$ we can always ensure that $c^2 = \pm 1$. There are therefore three distinct classes of real solutions to \eqref{mCYBE}: $c = i$, $c = 1$ and $c = 0$. These will be used in sections \ref{sec: type I}, \ref{sec: type II} and \ref{sec: type III} respectively to construct three different kinds of real deformations of integrable $\sigma$-models. We recall below the algebraic interpretation of equation \eqref{mCYBE} and its consequences in each of these three cases.

When $c \neq 0$, it will be convenient to introduce the shorthand notation $R^{\pm} \coloneqq R \pm c$. The notation does not depend explicitly on the parameter $c$ but this should not lead to confusion since it will usually be clear from the context what the value of $c$ is. In terms of this we can rewrite \eqref{mCYBE} in the form
\begin{equation} \label{mCYBE 2}
R^{\pm} \big( [R^{\pm} X, Y] + [X, R^{\mp} Y] \big) = \big[ R^{\pm} X, R^{\pm} Y \big].
\end{equation}
For later purposes it will be useful to rewrite this in tensorial notation. Let $R_{\1\2} \in \g \otimes \g$ denote the kernel of $R \in \text{End}\, \g$ with respect to the bilinear form on $\g$, with the property that $(RX)_{\1} = \langle R_{\1\2}, X_{\2} \rangle_{\2}$ for any $X \in \g$. Here we use the standard tensorial notation such as $X_{\1} \coloneqq X \otimes 1$ and $X_{\2} \coloneqq 1 \otimes X$. The kernel of the identity operator $\text{id} \in \text{End}\, \g$ is the split Casimir $C_{\1\2}$ satisfying $X_\1 = \langle C_{\1\2}, X_\2 \rangle_\2$ for any $X \in \g$. Define $R^\pm_{\1\2} \coloneqq R_{\1\2} + c \, C_{\1\2}$.
If $R$ is skew-symmetric with respect to the bilinear form on $\g$, which will always be the case in later sections, then we can rewrite \eqref{mCYBE 2} as
\begin{equation} \label{mCYBE tensor}
\big[ R^{\pm}_{\1\2}, R^{\pm}_{\1\3} \big] + \big[ R^{\pm}_{\1\2}, R^{\pm}_{\2\3} \big] +\big[ R^{\pm}_{\1\3}, R^{\pm}_{\2\3} \big] = 0.
\end{equation}
In the special case $c = 0$ we have the usual CYBE for $R$ in tensor form
\begin{equation} \label{CYBE tensor}
\big[ R_{\1\2}, R_{\1\3} \big] + \big[ R_{\1\2}, R_{\2\3} \big] +\big[ R_{\1\3}, R_{\2\3} \big] = 0.
\end{equation}

\subsection{The non-split case: $c = i$} \label{sec: non-split case}

Let $\g^{\CC} \coloneqq \g \otimes_{\mathbb{R}} \CC$ be the complexification of $\g$ which we regard as a real Lie algebra. The real form $\g$ then corresponds to the subalgebra of $\g^{\CC}$ fixed pointwise by an anti-linear involution $\tau : \g^{\CC} \to \g^{\CC}$, specifically $\tau(X \otimes u) = X \otimes \bar{u}$ for $X \in \g$ and $u \in \CC$. We denote by $G^\CC$ the connected and simply connected Lie group with Lie algebra $\g^\CC$.

Let $R \in \text{End}\, \g$ be a solution of mCYBE$(i)$ and consider the map
\begin{equation} \label{Rpm maps}
R^- = (R - i) : \g_R \longhookrightarrow \g^{\CC}.
\end{equation}
It is seen to be injective since for any $X \in \g$, if $RX - i X = 0$ then also $RX + iX = 0$ by applying the involution $\tau$ and hence $X = 0$.
Moreover, equation \eqref{mCYBE} says that \eqref{Rpm maps} is a homomorphism of real Lie algebras from $\g_R$ to $\g^{\CC}$.
We may therefore regard $\g_R$ as a subalgebra of $\g^{\CC}$ by identifying it with its image $R^-(\g_R)$ under the embedding $R^-$.

Consider the surjective map $\mu : \g^{\CC} \twoheadrightarrow \g$ defined by $\mu(X) = \frac{1}{2i} (X - \tau(X))$. Then the composition of linear maps
\begin{equation*}
\g_R \overset{R^-}\longhookrightarrow \g^{\CC} \overset{\mu}\longtwoheadrightarrow \g
\end{equation*}
is the identity on $\g$. In particular, the restriction of $\mu$ to $\text{im}\, R^-$, which we identify with $\g_R$ through the map $R^-$, is a bijection. Since the kernel of $\mu$ coincides with the real subalgebra $\g \subset \g^{\CC}$, we have a direct sum decomposition of vector spaces
\begin{equation} \label{ns decomp}
\g^{\CC} = \g \dotplus \g_R.
\end{equation}
where $\dotplus$ denotes the internal direct sum of subspaces. In fact, the converse is also true and we have the following.

\begin{proposition*}
The map $R \mapsto \g_R$ defines a one-to-one correspondence
between solutions of mCYBE$(i)$ on $\g$ and Lie subalgebras of $\g^{\CC}$ complementary to the real subalgebra $\g$, \emph{i.e.} $\p \subset \g^{\CC}$ such that $\g^{\CC} = \g \dotplus \p$.
\begin{proof}
Let $\p$ be a Lie subalgebra of $\g^{\CC}$ complementary to the real form $\g$. The corresponding solution $R_{\p} \in \text{End}\, \g$ of mCYBE$(i)$ is constructed as follows. By assumption we have the vector space decomposition $\g^{\CC} = \g \dotplus \p$. The above surjective map $\mu : \g^{\CC} \twoheadrightarrow \g$ whose kernel is $\g \subset \g^{\CC}$ therefore induces a linear isomorphism $\mu|_{\p} : \p \overset{\sim}\rightarrow \g$. In particular, any $X \in \g$ can be written uniquely in the form $X = - \mu(A) = \frac{i}{2} (A - \tau(A))$ for some $A \in \p$.
We now define
\begin{equation} \label{R def}
R_{\p} \big( \mbox{\small $\frac{i}{2}$} (A - \tau(A)) \big) \coloneqq \ha (A + \tau(A)).
\end{equation}
for $A \in \p$. It is straightforward to check that this satisfies \eqref{mCYBE} with $c = i$. Furthermore, the map $\p \mapsto R_{\p}$ is clearly the inverse of $R \mapsto \g_R$.
\end{proof}
\end{proposition*}

Consider the real non-degenerate bilinear form $\langle\!\langle \cdot, \cdot \rangle\!\rangle : \g^{\CC} \times \g^{\CC} \to \mathbb{R}$ on $\g^{\CC}$, viewed as a vector space over $\mathbb{R}$, defined for any $X, Y, X', Y' \in \g$ as
\begin{equation} \label{ip non-split}
\langle \! \langle X + i Y, X' + i Y' \rangle \! \rangle := \Im \langle X + i Y, X' + i Y' \rangle.
\end{equation}
A subspace $\mathfrak l \subset \g^\CC$ is said to be Lagrangian if it is maximal isotropic with respect to the bilinear form \eqref{ip non-split}, namely if $\langle \! \langle \mathfrak l, \mathfrak l \rangle \! \rangle = 0$ and $\mathfrak l$ is not properly contained in another subspace $\mathfrak m \subset \g^\CC$ with $\langle \! \langle \mathfrak m, \mathfrak m \rangle \! \rangle = 0$.

\begin{proposition*}
The subalgebra $\g \subset \g^{\CC}$ is Lagrangian. The complementary subalgebra $\g_R$ is also Lagrangian if and only if $R \in \text{End}\, \g$ is skew-symmetric.
\begin{proof}
It is clear that $\g$ is isotropic with respect to \eqref{ip non-split}. Maximality follows from the non-degeneracy of $\langle \cdot, \cdot \rangle$. As for $\g_R$, given any two elements $R^- X, R^- X' \in \g_R$ we have
\begin{equation*}
\langle \! \langle R^- X, R^- X' \rangle \! \rangle = \Im \langle R X - i X, R X' - i X' \rangle = - \langle RX, X' \rangle - \langle X, RX' \rangle.
\end{equation*}
Hence $\g_R$ is isotropic if and only if $R$ is skew-symmetric with respect to $\langle \cdot, \cdot \rangle$. The maximality then follows since $\g$ and $\g_R$ are complementary.
\end{proof}
\end{proposition*}

We shall refer to a skew-symmetric solution of mCYBE$(i)$ as a \emph{non-split} $R$-matrix. The above propositions show that there is a one-to-one correspondence between non-split $R$-matrices and Lagrangian subalgebras of $\g^\CC$ with respect to the bilinear form \eqref{ip non-split}, which are complementary to the real form $\g \subset \g^\CC$.

Given a non-split $R$-matrix, let $G_R$ denote the Lie subgroup of $G^\CC$ corresponding to the Lie subalgebra $\g_R \subset \g^\CC$. By the decomposition \eqref{ns decomp} of Lie algebras, we can factorise any elements $g \in G^\CC$ which lies in a small enough neighbourhood of the identity element into a product of elements in $G$ and $G_R$. That is, we can write $g = h h'$ for some $h \in G$ and $h' \in G_R$. Equivalently, by factorising instead the inverse element $g^{-1}$ in this way, we can also write $g = k' k$ for some $k' \in G_R$ and $k \in G$. However, in general, such a factorisation of elements of the Lie group $G^\CC$ fails to hold globally. Instead, if we choose representatives in $G^\CC$ for each double coset in $G \backslash G^\CC / G_R$ and denote by $\mathcal S \subset G^\CC$ the set of all such representatives, with $e \in \mathcal S$ by convention, then we may write $G^\CC$ as a disjoint union of cells \cite{AM94}
\begin{equation} \label{GC cell decomp}
G^\CC = \bigsqcup_{w \in \mathcal S} G w G_R.
\end{equation}
There is a neighbourhood of the identity in $G^\CC$ which belongs to the main cell $G G_R$, corresponding to the identity element $e \in \mathcal S$ in the above cell decomposition.
Similarly, applying the inversion map $G^\CC \to G^\CC$, $g \mapsto g^{-1}$ we also have the reverse decomposition
\begin{equation*}
G^\CC = \bigsqcup_{w \in \mathcal S} G_R w^{-1} G.
\end{equation*}
If \eqref{GC cell decomp} consists of a single cell, \emph{i.e.} $G^\CC = G G_R$ in which case we also have $G^\CC = G_R G$, then $G^\CC$ is known as the bicrossproduct $G \triangleright\!\!\triangleleft \, G_R$ of $G$ and $G_R$ \cite{Majid}.

\paragraph{Drinfel'd double.}
The complexified Lie algebra $\g^\CC$ is an example of a Drinfel'd double of $\g$ which we recall the definition of below.

Suppose that $R$ is a non-split $R$-matrix. Since $\g$ is finite dimensional, its dual $\g^\ast$ is isomorphic as a vector space to $\g_R \simeq \g$.
We can endow the latter with a coboundary $1$-cocycle $\delta : \g \to \g \wedge \g$ defined by the usual formula $\delta(X) \coloneqq [X_\1 + X_\2, R_{\1\2}]$ for any $X \in \g$. The dual map $\g^\ast \wedge \g^\ast \to \g^\ast$ defines a Lie bracket on $\g^{\ast}$ which coincides with the Lie bracket \eqref{R-bracket} on $\g_R$. Hence we have a natural isomorphism of Lie algebras $\g_R \cong \g^\ast$.

The Drinfel'd double of $\g$ is defined as
\begin{equation} \label{Drinfeld double} 
D(\g) \coloneqq \g \oplus \g^\ast.
\end{equation}
It is equipped with a natural symmetric bilinear form $Q : D(\g) \times D(\g) \to D(\g)$ defined by $Q(X + \alpha, Y + \beta) \coloneqq \beta(X) + \alpha(Y)$ for any $\alpha, \beta \in \g^\ast$ and $X, Y \in \g$.
Moreover, there exists a unique skew-symmetric bilinear operation $[\cdot, \cdot] : D(\g) \times D(\g) \to D(\g)$ such that
\begin{itemize}
  \item[$(i)$] it restricts to the Lie brackets on $\g$ and $\g^\ast$,
  \item[$(ii)$] the bilinear form $Q$ on $D(\g)$ is $\text{ad}$-invariant.
\end{itemize}

We have an isomorphism of Lie algebras
\begin{equation*}
\g^{\CC} \cong D(\g).
\end{equation*}
To see this, first note that $\g^\CC$ is isomorphic to $D(\g)$ as a vector space by virtue of the decomposition \eqref{ns decomp}. Next, the bilinear form \eqref{ip non-split} on $\g^\CC$ coincides with the above form $Q$ on $D(\g)$ since the subspaces $\g, \g_R \subset \g^\CC$ are both isotropic. Finally, the Lie bracket on $\g^\CC$ coincides with the above Lie bracket on $D(\g)$ since it satisfies both conditions $(i)$ and $(ii)$.
Indeed, the bilinear form \eqref{ip non-split} is clearly ad-invariant with respect to the Lie bracket of $\g^{\CC}$. Moreover, since $\g$ and $\g_R$ are both subalgebras of $\g^{\CC}$, the Lie bracket of $\g^{\CC}$ clearly restricts to the respective Lie brackets on these subalgebras.

\begin{example}
Let $\g^\CC$ be a complex semisimple Lie algebra and $\g \coloneqq \{ X \in \g^{\CC} \,|\, \tau(X) = X \}$
any real form specified by an anti-linear involution $\tau : \g^\CC \to \g^\CC$. Fix a Borel subalgebra $\b \subset \g^\CC$ such that $\g^\CC = \b + \tau(\b)$, namely $\b$ and $\tau(\b)$ are opposite Borel subalgebras. We introduce the Lie subalgebras $\h \coloneqq \b \cap \tau(\b)$, $\n \coloneqq [\b, \b]$ and $\a \coloneqq \{ h \in \h \,|\, \tau(h) = - h \}$. Regarding $\g^\CC$ as a Lie algebra over $\mathbb{R}$ we have the same vector space decomposition
\begin{equation} \label{gan decomp}
\g^{\CC} = \g \dotplus \a \dotplus \n.
\end{equation}
Indeed, since $\g^\CC = \n + \tau(\b)$ we may write any $x \in \g^\CC$ as $x = n + h + X$ where $n \in \n$, $X \in \tau(\n)$ and $h \in \tau(\h)$. On the other hand, we have
\begin{equation*}
X + h = \big( (X + \ha h) + \tau (X + \ha h) \big) + \ha \big(h - \tau(h)\big) - \tau(X) \in \g \dotplus \a \dotplus \n,
\end{equation*}
so that $x \in \g^\CC$ can be written as a sum in $\g \dotplus \a \dotplus \n$. Such a decomposition is clearly unique since the three subalgebras $\g$, $\a$ and $\n$ have pairwise trivial intersection. Moreover, it is straightforward to see that the subalgebras $\g$ and $\a \dotplus \n$ are both isotropic with respect to \eqref{ip non-split}. In particular, $\g_R = \a \dotplus \n$.
When $\g$ denotes the compact real form of $\g^\CC$, the splitting \eqref{gan decomp} is nothing but the Iwasawa decomposition. In this case we also have a decomposition at the level of Lie groups
\begin{equation*}
G^\CC = G A N = A N G,
\end{equation*}
where $A$ and $N$ are the subgroups of $G^\CC$ corresponding to the Lie subalgebras $\a$ and $\n$ of $\g^\CC$ respectively. In particular, we have $G_R = AN$. More generally, however, for other real forms $G$ we only have a cell decomposition as in \eqref{GC cell decomp}. See for instance \cite{Wolf, Aomoto}.
\end{example}

\subsection{The split case: $c = 1$} \label{sec: split case}

Define the real double of $\g$ as the direct sum of Lie algebras $\d \coloneqq \g \oplus \g$. The canonical projections $p_{\pm} : \d \rightarrow \g$ onto the first and second component respectively, are Lie algebra homomorphisms. Let $\psi : \d \to \d$ be the `flip' involution given by $\psi(X, Y) \coloneqq (Y, X)$. The diagonal subalgebra $\g^{\delta} \coloneqq \{ (X, X) \,|\, X \in \g \} \subset \d$ corresponds to the subspace of $\d$ fixed pointwise by $\psi$.

Given a solution $R \in \text{End}\, \g$ of mCYBE$(1)$ on $\g$, equation \eqref{mCYBE} says that the maps
\begin{equation} \label{Rtpm maps}
R^{\pm} = (R \pm 1) : \g_R \longrightarrow \g
\end{equation}
are both homomorphisms of real Lie algebras. Putting these maps together we obtain an embedding of real Lie algebras $\iota : \g_R \hookrightarrow \d$, $X \mapsto (R^+ X, R^- X)$ into $\d$. We will often identify $\g_R$ with its image under this embedding and regard $\g_R$ as a subalgebra of $\d$.

Consider the surjective map $\nu \coloneqq \ha (p_+ - p_-) : \d \twoheadrightarrow \g$. The composition of linear maps
\begin{equation*}
\g_R \overset{\iota}\longhookrightarrow \d \overset{\nu}\longtwoheadrightarrow \g
\end{equation*}
is the identity on $\g$. In particular, the restriction of $\nu$ to $\text{im} \, \iota$, which we identify with $\g_R$, is a bijection. Since the kernel of $\nu$ is $\g^{\delta}$, we have the direct sum decomposition of vector spaces
\begin{equation} \label{s decomp}
\d = \g^{\delta} \dotplus \g_R.
\end{equation}
We also have the converse statement.

\begin{proposition*}
The map $R \mapsto \g_R$ defines a one-to-one correspondence
between solutions of mCYBE$(1)$ on $\g$ and Lie subalgebras of $\d$ complementary to the diagonal subalgebra $\g^{\delta}$, \emph{i.e.} $\p \subset \d$ such that $\d = \g^{\delta} \dotplus \p$.
\begin{proof}
Let $\p$ be a subalgebra of $\d$ complementary to the diagonal $\g^{\delta}$. We will define a corresponding solution $R_\p$ of mCYBE$(1)$. Since by assumption we have $\d = \g^{\delta} \dotplus \p$, the above surjective map $\nu : \d \twoheadrightarrow \g$ induces a linear isomorphism $\nu|_{\p} = \ha (r_+ - r_-)$ between $\p$ and $\g$, where $r_{\pm} \coloneqq p_{\pm}|_{\p}$.

Letting $R_\p^{\pm} \coloneqq r_{\pm} \circ \nu|^{-1}_{\p} \in \text{End } \g$ we have $\ha (R_\p^+ - R_\p^-) = 1$. Define $R_\p \coloneqq \ha (R_\p^+ + R_\p^-)$ so that $R_\p^{\pm} = R_\p \pm 1$. Since $p_\pm : \d \rightarrow \g$ are homomorphisms of Lie algebras, so are their restrictions $r_{\pm} : \p \rightarrow \g$ to the Lie subalgebra $\p \subset \d$. That is, $[r_{\pm} A, r_{\pm} B] = r_{\pm} \big( [A, B]_{\d} \big)$, for any $A, B \in \p$. Hence, writing $r_{\pm} = R_\p^{\pm} \circ \nu|_{\p}$ we have
\begin{align*}
\big[ R_\p^{\pm} (\nu A), R_\p^{\pm} (\nu B) \big] &= R^{\pm} \big( \nu [A, B]_{\d} \big) = R_\p^{\pm} \big( \ha [p_+ A, p_+ B] - \ha [p_- A, p_- B] \big)\\
&= R_\p^{\pm} \left( \ha \big[ R_\p^+ (\nu A), R_\p^+ (\nu B) \big] - \ha \big[ R_\p^- (\nu A), R_\p^- (\nu B) \big] \right),
\end{align*}
where in the second equality we used the definition of $\nu$ and the fact that $p_{\pm}$ are both homomorphisms of Lie algebras.

Let $X, Y \in \g$ be arbitrary. Since $\nu|_{\p}$ is a bijection from $\p$ to $\g$, there exists $A, B \in \p$ such that $X = \nu A \in \g$ and $Y = \nu B \in \g$. It then follows from above that
\begin{equation*}
\big[ R_\p^{\pm} X, R_\p^{\pm} Y \big] = R_\p^{\pm} \big( [R_\p X, Y] + [X, R_\p Y] \big),
\end{equation*}
which is precisely mCYBE$(1)$ in the form \eqref{mCYBE}, with $c = 1$, for $R_\p \in \text{End } \g$.
The maps $\p \mapsto R_\p$ and $R \mapsto \g_R$ are seen to be inverses of each other.
\end{proof}
\end{proposition*}

We introduce the real non-degenerate bilinear form $\langle \! \langle \cdot, \cdot \rangle \! \rangle : \d \times \d \to \mathbb{R}$ on the double $\d$, defined for any $X, Y, X', Y' \in \g$ as
\begin{equation} \label{ip split}
\langle \! \langle (X, Y), (X', Y') \rangle \! \rangle := \langle X, X' \rangle - \langle Y, Y' \rangle.
\end{equation}
As in the non-split case, a subspace $\mathfrak l \subset \d$ is called Lagrangian if it is maximal isotropic with respect to the bilinear form \eqref{ip split}.

\begin{proposition*}
The diagonal $\g^{\delta}$ is a Lagrangian subalgebra of $\d$. The complementary subalgebra $\g_R$ is also Lagrangian if and only if $R \in \text{End}\, \g$ is skew-symmetric.
\begin{proof}
The isotropy of $\g^{\delta}$ is evident and the maximality follows from the non-degeneracy of $\langle \cdot, \cdot \rangle$. For $\g_R$, the condition for isotropy follows from the relation
\begin{equation*}
\langle \! \langle \iota X, \iota Y \rangle \! \rangle = \langle R^+ X, R^+ Y \rangle - \langle R^- X, R^- Y \rangle = 2 \langle R X, Y \rangle + 2 \langle X, R Y \rangle.
\end{equation*}
for any $\iota X, \iota Y \in \g_R$. Maximality then follows since $\g^{\delta}$ and $\g_R$ are complementary.
\end{proof}
\end{proposition*}

We will refer to a skew-symmetric solution of mCYBE$(1)$ as a \emph{split} $R$-matrix. By the above two propositions we therefore have a one-to-one correspondence between split $R$-matrices and Lagrangian subalgebras of $\d$ with respect to the bilinear form \eqref{ip split}, complementary to the diagonal subalgebra $\g^\delta \subset \d$.

As in the non-split case, the decomposition \eqref{s decomp} of the real double Lie algebra $\d$ of $\g$ into complementary subalgebras can be extended to the group as follows. Consider the real Lie group $D \coloneqq G \times G$ with Lie algebra $\d$ and let $G^\delta \coloneqq \{ (x, x) \,|\, x \in G \} \subset D$ be the subgroup corresponding to the diagonal subalgebra $\g^\delta \subset \d$. Given a split $R$-matrix, let $G_R$ denote the subgroup of $D$ associated with the Lie subalgebra $\g_R$ of $\d$. We can factorise an element $d \in D$ lying in the vicinity of the identity in $D$ as $d = h g$ for some $g \in G_R$ and $h \in G^\delta$. However, this factorisation may fail to hold globally in general. Instead we have the cell decomposition
\begin{equation} \label{D cell decomp}
D = \bigsqcup_{w \in \mathcal S} G^\delta w G_R,
\end{equation}
where $\mathcal S \subset D$ is a set of representatives for each double coset in $G^\delta \backslash D / G_R$. We choose $e \in \mathcal S$ corresponding to the main cell $G^\delta G_R \subset D$ which contains a neighbourhood of the identity in $D$. Applying the inversion map we also obtain the decomposition in the reverse order
\begin{equation} \label{D cell decomp reverse}
D = \bigsqcup_{w \in \mathcal S} G_R w^{-1} G^\delta.
\end{equation}

\paragraph{Drinfel'd double.} The real double $\d = \g \oplus \g$ is also an example of a Drinfel'd double of $\g$ but corresponding to a different $R$-matrix, namely a split one.

Suppose $R$ is a split $R$-matrix. By the exact same reasoning as in the non-split case, we can endow $\g^\ast$ with a Lie bracket $[\cdot, \cdot]_\ast$ obtained as the dual map to a coboundary $1$-cocycle defined using the given $R$-matrix. In particular, we have a natural isomorphism of Lie algebras $\g_R \cong \g^\ast$. The corresponding Drinfel'd double is given as a vector space by the same expression \eqref{Drinfeld double}, it is equipped with the same bilinear form $Q$ and its Lie bracket is also defined by conditions $(i)$ and $(ii)$. However, since the Lie bracket $[\cdot, \cdot]_\ast$ on $\g^\ast$ is different, the resulting Drinfel'd double $D(\g)$ is different. We have, in the present case, the natural isomorphism of Lie algebras
\begin{equation*}
\d \cong D(\g).
\end{equation*}

\begin{example}
Let $\g^\CC = \h \oplus \bigoplus_{\alpha \in \Phi} \g_\alpha$ be a complex semisimple Lie algebra with Cartan-Weyl basis $h_j \in \h$ for $j = 1, \ldots, \text{rk}\, \g$ and $e_\alpha \in \g_\alpha$ for $\alpha \in \Phi$. Then the split real form $\g$ is defined as the real span of the Cartan-Weyl basis. Introduce the nilpotent subalgebras $\n_{\pm}^{\mathbb{R}} \coloneqq \bigoplus_{\alpha \in \Phi^+} \mathbb{R} e_{\pm \alpha}$ and the Cartan subalgebra $\h^{\mathbb{R}} \coloneqq \bigoplus_{j=1}^{\text{rk}\, \g} \mathbb{R} h_j$. We have the Gauss decomposition of the split real form $\g$ given by the direct sum of vector spaces
\begin{equation} \label{Gauss decomp}
\g = \n^{\mathbb{R}}_- \dotplus \h^{\mathbb{R}} \dotplus \n_+^{\mathbb{R}}.
\end{equation}
Denote the projections relative to this decomposition as $P_{\pm} : \g \to \n_{\pm}^{\mathbb{R}}$ and $P_0 : \g \to \h^{\mathbb{R}}$. A skew-symmetric solution of mCYBE$(1)$ on $\g$ is given explicitly by $R \coloneqq P_+ - P_-$ and we have the corresponding decomposition of the real double
\begin{equation*}
\d = \g^\delta \dotplus \big\{ ( \ha P_0 X + P_+ X, - \ha P_0 X - P_- X) \in \d \,\big|\, X \in \g \big\}.
\end{equation*}
\end{example}

\subsection{The CYBE case: $c = 0$} \label{sec: CYBE case}

When $c = 0$, the equation \eqref{mCYBEc} reduces to the usual classical Yang-Baxter equation (CYBE). Therefore we denote mCYBE$(0)$ simply by CYBE. Although this case will not play a big role in our analysis, we will briefly describe the algebraic interpretation of this equation analogous to the non-split and split cases above.

Algebraically, equation \eqref{mCYBEc} says that the map
\begin{equation} \label{R map}
R : \g_R \longrightarrow \g
\end{equation}
is itself a real Lie algebra homomorphism. In particular, $\k \coloneqq \text{im}\, R$ is a Lie subalgebra of $\g$.
Restricting the codomain of \eqref{R map} to the image $\k$, we obtain a surjection which we denote also $R : \g_R \twoheadrightarrow \k$. The latter therefore has a right inverse $R^{-1} : \k \to \g_R$.
Consider the bilinear pairing $\omega : \k \wedge \k \to \mathbb{R}$ defined by $\omega(X, Y) \coloneqq \langle R^{-1} X, Y \rangle$ for any $X, Y \in \k$.
By using the skew-symmetry of the $R$-matrix we find that $\omega$ is skew-symmetric, namely
\begin{equation*}
\omega(X, Y) = - \omega(Y, X)
\end{equation*}
for all $X, Y \in \k$. Moreover, it follows again using the skew-symmetry of $R$ together with equation \eqref{mCYBE} that $\omega$ defines a 2-cocycle on $\k$, namely
\begin{equation*}
\omega \big( [X, Y], Z \big) + \omega \big( [Y, Z], X \big) + \omega \big( [Z, X], Y \big) = 0,
\end{equation*}
for all $X, Y, Z \in \k$. This makes $\k$ into a quasi-Frobenius Lie algebra. In general we have the following characterisation of solutions to CYBE.

\begin{proposition*}
The map $R \mapsto \g_R$ establishes a one-to-one correspondence
between skew-symmetric solutions of CYBE on $\g$ and quasi-Frobenius Lie subalgebras of $\g$.
\end{proposition*}

\section{Constructing deformations} \label{sec: def}

Consider the loop group $\Loop G \coloneqq C^\infty(S^1, G)$ consisting of smooth maps from the circle to the real Lie group $G$. Its Lie algebra is the space $\Loop \g \coloneqq C^{\infty}(S^1, \g)$ of smooth loops into $\g = \text{Lie}(G)$. Using the bilinear form $\langle \cdot, \cdot \rangle : \g \times \g \to \mathbb{R}$ on $\g$ we equip the loop algebra $\Loop \g$ with a non-degenerate bilinear form $( \cdot | \cdot ) : \Loop \g \times \Loop \g \to \mathbb{R}$ defined as
\begin{equation} \label{Lg ip}
( x | y ) \coloneqq \int_{S^1} \langle x(\theta), y(\theta) \rangle d\theta
\end{equation}
for any $x, y \in \Loop \g$. The linear map $\Loop \g \to (\Loop \g)'$ given by $x \mapsto (x|\cdot)$ identifies $\Loop \g
$ with a subspace of the algebraic dual $(\Loop \g)'$ called the smooth dual $\Loop \g^\ast$.
It will also be useful to extend \eqref{Lg ip} to the complexified loop algebra $\Loop \g^\CC$ by $\CC$-linearity. Namely, we define $( \cdot | \cdot ) : \Loop \g^\CC \times \Loop \g^\CC \to \CC$ by the same expression as \eqref{Lg ip} but for $x, y \in \Loop \g^\CC$.

\paragraph{Phase space.}
We are interested in describing integrable field theories whose phase space is the cotangent bundle $T^\ast \Loop G$ of $\Loop G$. Using the global (left) trivialisation we can identify $T^\ast \Loop G$ with the Cartesian product $\Loop G \times (\Loop \g)'$. Strictly speaking we shall work on the subspace $\Loop G \times \Loop \g^\ast$, which by abuse of notation we will keep calling $T^\ast \Loop G$. This way, $T^\ast \Loop G$ is parameterised by a pair of fields $g \in \Loop G$ and $X \in \Loop \g^\ast \simeq \Loop \g$. In terms of these, the canonical Poisson brackets on the cotangent bundle $T^\ast \Loop G$ read
\begin{subequations} \label{PB g X}
\begin{align}
\label{gg PB}
\{ g_{\1}(\theta), g_{\2}(\theta') \}_0 &= 0,\\
\label{Xg PB}
\{ X_{\1}(\theta), g_{\2}(\theta') \}_0 &= g_{\2}(\theta) C_{\1\2} \delta_{\theta \theta'},\\
\label{XX PB}
\{ X_{\1}(\theta), X_{\2}(\theta') \}_0 &= - [C_{\1\2}, X_{\2}(\theta)] \delta_{\theta \theta'}.
\end{align}
\end{subequations}
We will also be interested in describing WZW-type models, whose dynamics cannot be expressed in terms of the above Poisson structure. In this case, a suitable phase space is given by the cotangent bundle $T^\ast \Loop G$ but equipped with a modified Poisson structure. Specifically, if we label the fields parameterising the global (left) trivialisation of $T^\ast \Loop G$ as $\F \in \Loop G$ and $\J \in \Loop \g$, then the relevant one-parameter family of Poisson brackets on $T^\ast \Loop G$ is given by \cite{HK95}
\begin{subequations} \label{PB F J}
\begin{align}
\label{FF PB}
\{ \F_{\1}(\theta), \F_{\2}(\theta') \}_\kappa &= 0,\\
\label{JpF PB}
\{ \J_{\1}(\theta), \F_{\2}(\theta') \}_\kappa &= \F_{\2}(\theta) C_{\1\2} \delta_{\theta \theta'},\\
\label{JJ PB}
\{ \J_{\1}(\theta), \J_{\2}(\theta') \}_\kappa &= - [C_{\1\2}, \J_{\2}(\theta)] \delta_{\theta \theta'} - \kappa \, C_{\1\2} \delta'_{\theta \theta'},
\end{align}
\end{subequations}
where $\kappa \in \RR$ is a non-zero real parameter.

\paragraph{Lax matrix.}
The requirement of integrability means that there should exist a rational function of the spectral parameter $\lambda \in \CC$ taking values in the complexified loop algebra $\Loop \g^\CC \coloneqq C^{\infty}(S^1, \g^\CC)$, called the Lax matrix $\L(\lambda, \theta)$, whose Poisson bracket with itself is of the general non-ultralocal type \cite{Maillet86}. In the notation of the present paper we write the latter as
\begin{align} \label{r-s algebra}
\big\{ \L_{\1}(\lambda, \theta), \L_{\2}(\mu, \theta') \big\} &= \big[\R_{\1\2}(\lambda, \mu), \L_{\1}(\lambda, \theta)\big] \delta_{\theta \theta'} - \big[\R_{\2\1}(\mu, \lambda), \L_{\2}(\mu, \theta)\big] \delta_{\theta \theta'} \notag\\
&\qquad\qquad\qquad\qquad\qquad - \big( \R_{\1\2}(\lambda, \mu) + \R_{\2\1}(\mu, \lambda) \big) \delta'_{\theta \theta'}.
\end{align}
Here the $\R$-matrix $\R_{\1\2}(\lambda, \mu)$ is a $\g^\CC \otimes \g^\CC$-valued rational function of the spectral parameters $\lambda$ and $\mu$ satisfying the classical Yang-Baxter equation, which in tensorial notation reads
\begin{equation} \label{CYBE tensor cal R}
[\R_{\1\2}(\lambda, \mu), \R_{\1\3}(\lambda, \nu)] +
[\R_{\1\2}(\lambda, \mu), \R_{\2\3}(\mu, \nu)] +
[\R_{\3\2}(\nu, \mu), \R_{\1\3}(\lambda, \nu)] = 0.
\end{equation}
As usual, for any element $\mathcal{O}_{\1\2} \in \g^\CC \otimes \g^\CC$, we define $\mathcal{O}_{\2\1} \coloneqq P(\mathcal{O}_{\1\2})$ where $P : \g^\CC \otimes \g^\CC \to \g^\CC \otimes \g^\CC$ is the permutation operator given by $P(a \otimes b) = b \otimes a$.

\paragraph{Twist function.}
We shall assume that the $\R$-matrix is of the form
\begin{equation} \label{scR-matrix def}
\R_{\1\2}(\lambda, \mu) = \R^0_{\1\2}(\lambda, \mu) \varphi(\mu)^{-1},
\end{equation}
where $\R^0_{\1\2}(\lambda, \mu)$ is another $\g^\CC \otimes \g^\CC$-valued rational function of $\lambda$ and $\mu$ also satisfying \eqref{CYBE tensor cal R} but distinguished by its leading order behaviour in the limit $\lambda \to \mu$ being of the form
\begin{equation} \label{scR0-matrix}
\R^0_{\1\2}(\lambda, \mu) = \frac{C_{\1\2}}{\mu - \lambda} + \mathcal{O}\big( (\lambda - \mu)^0 \big).
\end{equation}
The rational function $\varphi$ defined through \eqref{scR-matrix def} is called the twist function and will play a central role in our construction.

Our strategy for constructing integrable field theories on $T^\ast \Loop G$ is as follows. Starting from a given integrable field theory of the above type, we shall deform it by modifying only its twist function and keeping, in particular, the same Lax matrix $\L(\lambda, \theta)$. In this way the resulting field theory will be manifestly integrable. The main difficulty will be to identify a suitable model on the cotangent bundle $T^\ast \Loop G$ which has the new deformed twist function. This requires extracting from the Lax matrix $\L(\lambda, \theta)$ a pair of fields in $\Loop G$ and $\Loop \g$ whose Poisson brackets are given either by \eqref{PB g X} or by \eqref{PB F J}.

More precisely, we will start from a model on $T^\ast \Loop G$ whose twist function $\varphi(\lambda)$ has a \emph{double} pole at some point $\lambda_0 \in \RR$. The deformed theory will then simply be defined by exchanging this double pole at $\lambda_0$ for a pair of \emph{simple} poles. In order to extract the field content of the deformed theory from these simple poles we shall make essential use of the following result.

\begin{proposition} \label{prop: KM currents}
At any simple pole $z$ of the twist function $\varphi(\lambda)$, the value $\L(z, \cdot)$ of the Lax matrix defines a Kac-Moody current of level $\kappa \coloneqq \res_z \varphi(\lambda) d\lambda$. Specifically, we have
\begin{subequations} \label{KM algebra}
\begin{equation} \label{KM algebra a}
\kappa \big\{ \L_{\1}(z, \theta), \L_{\2}(z, \theta') \big\} = - \big[ C_{\1\2}, \L_{\2}(z, \theta) \big] \delta_{\theta \theta'} - C_{\1\2} \delta'_{\theta \theta'}.
\end{equation}
Moreover, if $z \neq z'$ are any two distinct simple poles of $\varphi(\lambda)$, then the corresponding Kac-Moody currents Poisson commute
\begin{equation} \label{KM algebra b}
\big\{ \L_{\1}(z, \theta), \L_{\2}(z', \theta') \big\} = 0.
\end{equation}
\end{subequations}
\begin{proof}
Expanding the Lax matrix algebra \eqref{r-s algebra} in small $\lambda - \mu$ and using the explicit form \eqref{scR-matrix def} of the $\R$-matrix with \eqref{scR0-matrix} we find
\begin{align*}
\big\{ \L_{\1}(\lambda, \theta), \L_{\2}(\mu, \theta') \big\} &= \varphi(\mu)^{-1} \left( \frac{1}{\mu - \lambda} [ C_{\1\2}, \L_\1(\mu, \theta) + \L_\2(\mu, \theta)] \delta_{\theta \theta'} + \mathcal O \big( (\lambda - \mu)^0 \big) \right)\\
&\qquad\quad + \frac{d \varphi(\mu)^{-1}}{d\mu} \big( - [C_{\1\2}, \L_\2(\mu, \theta)] \delta_{\theta \theta'} - C_{\1\2} \delta'_{\theta \theta'} \big) + \mathcal O(\lambda - \mu).
\end{align*}
The singular term in $(\lambda - \mu)^{-1}$ on the right hand side above vanishes by the $\g^\CC$-invariance of the split Casimir, namely $[C_{\1\2}, X_\1 + X_\2] = 0$ for any $X \in \g^\CC$.
The first result \eqref{KM algebra a} now follows by setting $\lambda = \mu = z$ and using the fact that for a function $\varphi$ with a simple pole at $z$ we have
\begin{equation*}
\left( \left. \frac{d \varphi(\mu)^{-1}}{d\mu}\right|_{\mu = z} \right)^{-1} = \res_z \varphi(\lambda) d\lambda.
\end{equation*}
The second result \eqref{KM algebra b} is immediate from \eqref{r-s algebra} after setting $\lambda = z$, $\mu = z' \neq z$.
\end{proof}
\end{proposition}

For concreteness we will illustrate the construction of deformations in the case of symmetric space $\sigma$-models, but the same procedure described below applies more generally to other integrable field theories of the above type, such as principal chiral models and semi-symmetric space $\sigma$-models.

\subsection{The undeformed model}

The model with phase space $T^\ast \Loop G$ and Poisson brackets \eqref{PB g X} which we shall start with is the symmetric space $\sigma$-model on some quotient $G/G^{(0)}$ of $G$. To define its Hamiltonian and integrable structure, it is convenient first to pass to the quotient $G \backslash T^\ast \Loop G$ by the natural left Hamiltonian action of the subgroup of constant loops $G$. Specifically, we can identify the quotient space $G \backslash T^\ast \Loop G$ with $\Loop \g \times \Loop \g^\ast \simeq \Loop \g \times \Loop \g$ by introducing the pair of fields $A, \Pi \in \Loop \g$ defined as
\begin{equation} \label{APi def}
A \coloneqq - \partial_{\theta} g\, g^{-1}, \qquad
\Pi \coloneqq - g X g^{-1}.
\end{equation}
Their Poisson brackets follow from \eqref{PB g X} and take the form
\begin{subequations} \label{PB A Pi}
\begin{align}
\big\{ A_{\1}(\theta), A_{\2}(\theta') \big\} &= 0,\\
\big\{ A_{\1}(\theta), \Pi_{\2}(\theta') \big\} &= - \big[ C_{\1\2}, A_{\2}(\theta) \big] \delta_{\theta \theta'} - C_{\1\2} \delta'_{\theta \theta'},\\
\big\{ \Pi_{\1}(\theta), \Pi_{\2}(\theta') \big\} &= - \big[ C_{\1\2}, \Pi_{\2}(\theta) \big] \delta_{\theta \theta'}.
\end{align}
\end{subequations}
Now let $\sigma : \g \to \g$ be an automorphism of $\g$ of order $2$ and denote by $\g = \g^{(0)} \oplus \g^{(1)}$ the decomposition of $\g$ into eigenspaces of $\sigma$. In particular $\g^{(0)} \subset \g$ is the subalgebra fixed pointwise by $\sigma$. Let $G^{(0)}$ be the corresponding subgroup of $G$. Denoting by $A^{(i)}$, $\Pi^{(i)}$ for $i \in \mathbb{Z}_2$ the components of the fields $A, \Pi \in \Loop \g$ along $\g^{(i)}$, the Hamiltonian can now be defined, with the help of the bilinear form \eqref{Lg ip}, as
\begin{equation} \label{Ham A Pi}
\mathcal H = \ha \big( A^{(1)} \big| A^{(1)} \big) + \ha \big( \Pi^{(1)} \big| \Pi^{(1)} \big) + \big( A^{(0)} \big| \Pi^{(0)} \big).
\end{equation}

\paragraph{Lax matrix.}

The Lax matrix $\L(\lambda, \theta)$ can also be expressed in terms of the graded components $A^{(i)}$, $\Pi^{(i)}$ of the fields. It is given explicitly by
\begin{equation} \label{Lax matrix sym}
\L(\lambda, \cdot) = A^{(0)} + \ha (\lambda + \lambda^{-1}) A^{(1)} + \ha (\lambda^2 - 1) \Pi^{(0)} + \ha (\lambda - \lambda^{-1}) \Pi^{(1)}.
\end{equation}
Recall the anti-linear involution $\tau$ of $\g^\CC$ of which $\g$ is the fixed point subalgebra. We extend $\tau$ to the loop algebra $\Loop \g^\CC$ by defining $\tau : \Loop \g^\CC \to \Loop \g^\CC$ as $\tau(X) (\theta) \coloneqq \tau(X(\theta))$.
Since both fields $A$ and $\Pi$ are valued in the real Lie algebra $\g$ they satisfy $\tau(A) = A$ and $\tau(\Pi) = \Pi$ from which it follows that the Lax matrix satisfies the reality condition
\begin{equation} \label{reality Lax}
\tau\big( \L(\lambda, \theta) \big) = \L(\overline{\lambda}, \theta).
\end{equation}
Furthermore, the behaviour of the Lax matrix under the automorphism $\sigma$ follows from the properties of the graded components $A^{(i)}, \Pi^{(i)}$ of the fields $A, \Pi$ under the action of $\sigma$, namely $\sigma(A^{(i)}) = (-1)^i A^{(i)}$ and $\sigma(\Pi^{(i)}) = (-1)^i \Pi^{(i)}$. Explicitly, we have
\begin{equation} \label{sigma Lax}
\sigma\big( \L(\lambda, \theta) \big) = \L(-\lambda, \theta).
\end{equation}

The Poisson bracket of the Lax matrix \eqref{Lax matrix sym} with itself is found to be of the form \eqref{r-s algebra}. In particular, the $\R$-matrix is given by
\begin{equation} \label{scR-matrix}
\R_{\1\2}(\lambda, \mu) = 2 \frac{\lambda C^{(00)}_{\1\2} + \mu C^{(11)}_{\1\2}}{\mu^2 - \lambda^2} \varphi(\mu)^{-1},
\end{equation}
where $C^{(ii)}_{\1\2}$ for $i \in \mathbb{Z}_2$ are the graded components of the split Casimir of $\g$ and the twist function reads \cite{Sevostyanov, Vicedo:2010qd}
\begin{equation} \label{coset twist}
\varphi(\lambda) \coloneqq \frac{4 \lambda}{(1 - \lambda^2)^2}.
\end{equation}
Note that on general grounds, for the algebra \eqref{r-s algebra} with $\R$-matrix \eqref{scR-matrix} to be preserved under the commuting automorphisms $\tau$ and $\sigma$, the twist function $\varphi(\lambda)$ should be both real and odd, namely
\begin{equation} \label{twist real odd}
\overline{\varphi(\lambda)} = \varphi(\overline{\lambda}), \qquad
\varphi(- \lambda) = - \varphi(\lambda).
\end{equation}

As described above, the pertinent feature of the twist function \eqref{coset twist} which we shall exploit to construct deformations is the fact that it has double poles along the real axis, specifically at $\lambda = \pm 1$. In fact, since the twist function is necessarily odd, only one of these poles, say $\lambda = 1$, will be relevant.
More generally, the construction of deformations presented below for integrable field theories with phase space $T^\ast \Loop G$ can be applied to any model whose twist function $\varphi(\lambda)$ has a double pole on the real axis.

\subsection{Reconstructing the undeformed model}

In the previous subsection we started out with a given model defined by its Hamiltonian \eqref{Ham A Pi} on the cotangent bundle $T^\ast \Loop G$ which is equipped with the Poisson brackets \eqref{PB g X}. We then described its integrable structure by introducing the Lax matrix \eqref{Lax matrix sym}. In the next sections, however, since the model will be unknown to begin with we will have to do the reverse. That is, starting from the same Lax matrix \eqref{Lax matrix sym} as above but together with a new twist function $\varphi(\lambda)$, we will have to extract from this data a set of fields in $\Loop G$ and $\Loop \g$ parameterising $T^\ast \Loop G$ with Poisson brackets of the form \eqref{PB g X} or \eqref{PB F J} as well as define the Hamiltonian of our model. Before doing this in the deformed case, we will end this section by describing the procedure in the case at hand where the twist function is still given by \eqref{coset twist}.

\paragraph{Poisson brackets.} The fields $A, \Pi \in \Loop \g$ can be extracted directly from the behaviour of the Lax matrix near the double pole $\lambda = 1$ of the twist function.
Indeed, consider the first two terms in the expansion of the Lax matrix there, which we write as
\begin{equation*}
\L(\lambda, \theta) = \L(1, \theta) + \L'(1, \theta) (\lambda - 1) + \mathcal{O}\big( (\lambda - 1)^2 \big).
\end{equation*}
If we were to use the explicit form \eqref{Lax matrix sym} of the Lax matrix in terms of the phase space fields then we would find that $\L(1, \theta) = A(\theta)$ and $\L'(1, \theta) = \Pi(\theta)$. However, in order to emphasise how the different ingredients are obtained from the Lax matrix, throughout the remainder of this section we will keep expressing everything in terms of $\L(\lambda, \theta)$.

By performing a similar calculation to that used in the proof of Proposition \ref{prop: KM currents} and using the explicit form \eqref{coset twist} of the twist function for the model at hand, the Poisson bracket \eqref{r-s algebra} of the Lax matrix leads to the following brackets
\begin{subequations} \label{LL algebra undef}
\begin{align}
\label{LL algebra undef a} \big\{ \L_{\1}(1, \theta), \L_{\2}(1, \theta') \big\} &= 0,\\
\label{LL algebra undef b} \big\{ \L_{\1}(1, \theta), \L'_{\2}(1, \theta') \big\} &= - \big[ C_{\1\2}, \L_{\2}(1, \theta) \big] \delta_{\theta \theta'} - C_{\1\2} \delta'_{\theta \theta'},\\
\label{LL algebra undef c} \big\{ \L'_{\1}(1, \theta), \L'_{\2}(1, \theta') \big\} &= - \big[ C_{\1\2}, \L'_{\2}(1, \theta) \big] \delta_{\theta \theta'}.
\end{align}
\end{subequations}
In light of the above, these are nothing but the canonical Poisson brackets \eqref{PB A Pi} of the fields $A, \Pi \in \Loop \g$.

\paragraph{Lift to the cotangent bundle.} Next, we wish to extract a pair of fields $g \in \Loop G$ and $X \in \Loop \g$ satisfying the canonical Poisson brackets \eqref{PB g X} on the cotangent bundle $T^\ast \Loop G$.

Consider the so called extended solution $\Psi(\lambda, \theta)$ which by definition satisfies
\begin{equation} \label{ext sol}
- \partial_{\theta} \Psi(\lambda, \theta) \Psi(\lambda, \theta)^{-1} = \L(\lambda, \theta),
\end{equation}
and with the initial condition at $\theta = 0$ having the reality property $\tau\big( \Psi(\lambda, 0) \big) = \Psi(\overline{\lambda}, 0)$.
Using the reality condition on the Lax matrix \eqref{reality Lax} and the fact that $\tau$ is a homomorphism, it follows that
\begin{equation} \label{reality Psi}
\tau\big( \Psi(\lambda, \theta) \big) = \Psi(\overline{\lambda}, \theta).
\end{equation}
We define the field $g \in \Loop G$ to be the value of the extended solution at the double pole of the twist function, namely
\begin{equation} \label{g undef}
g \coloneqq \Psi(1, \cdot).
\end{equation}
In particular, by the reality condition \eqref{reality Psi} it takes values in $G$. The field $X \in \Loop \g$ is then defined by
\begin{equation} \label{X undef}
X \coloneqq - g^{-1} \L'(1,\cdot) g.
\end{equation}
With the fields $g$ and $X$ so defined, the Poisson brackets \eqref{LL algebra undef} are seen to follow from \eqref{PB g X}, as required.

\paragraph{Hamiltonian.} Finally, the Hamiltonian of the model is a quadratic expression of the phase space fields $A$ and $\Pi$, which can be extracted from the poles $\lambda = 0$ and $\lambda = \infty$ of the Lax matrix. Indeed, the Hamiltonian can be defined as \cite{Vicedo:2009sn}
\begin{equation} \label{Ham def}
\mathcal H \coloneqq \qa (\res_{\lambda = 0} - \res_{\lambda = \infty}) \big( \L(\lambda, \cdot) \big| \L(\lambda, \cdot) \big) \varphi(\lambda) d\lambda.
\end{equation}
Evaluating this using the explicit form of the Lax matrix \eqref{Lax matrix sym} and the twist function \eqref{coset twist} we recover the expression for the Hamiltonian in terms of the phase space fields as given in \eqref{Ham A Pi}.

\section{The complex branch} \label{sec: type I}

\begin{figure}[h]
\centering
\def\svgwidth{60mm}
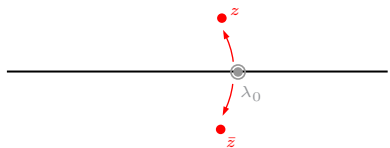
\caption{The complex branch.}
\label{fig: poles coset I}
\end{figure}

One way to deform the twist function $\varphi(\lambda)$ given in \eqref{coset twist} with a double pole at $\lambda = 1$, while preserving the reality condition in \eqref{twist real odd}, is to replace the double pole by a pair of complex conjugate simple poles $z, \bar{z}$ with $z \neq \bar{z}$, as depicted in Figure \ref{fig: poles coset I}.
In particular, we do not alter its zeroes. Specifically, we define a new twist function as
\begin{equation} \label{twist I}
\varphi_{\sf c}(\lambda) \coloneqq \frac{4 \lambda}{(\lambda^2 - z^2)(\lambda^2 - \bar{z}^2)}.
\end{equation}
We refer to this as the complex branch. Correspondingly, we also define a new $\R$--matrix by the same expression as in \eqref{scR-matrix} but using the new twist function, that is we set
\begin{equation*}
\R^{\sf c}_{\1\2}(\lambda, \mu) \coloneqq 2 \frac{\lambda C^{(00)}_{\1\2} + \mu C^{(11)}_{\1\2}}{\mu^2 - \lambda^2} \varphi_{\sf c}(\mu)^{-1}.
\end{equation*}
The deformed Poisson algebra of the Lax matrix then takes the exact same form as in \eqref{r-s algebra} but with the above $\R$-matrix
\begin{align} \label{r-s algebra I}
\big\{ \L_{\1}(\lambda, \theta), \L_{\2}(\mu, \theta') \big\}_{\sf c} &= \big[\R^{\sf c}_{\1\2}(\lambda, \mu), \L_{\1}(\lambda, \theta)\big] \delta_{\theta \theta'} - \big[\R^{\sf c}_{\2\1}(\mu, \lambda), \L_{\2}(\mu, \theta)\big] \delta_{\theta \theta'} \notag\\
&\qquad\qquad\qquad\qquad\qquad - \big( \R^{\sf c}_{\1\2}(\lambda, \mu) + \R^{\sf c}_{\2\1}(\mu, \lambda) \big) \delta'_{\theta \theta'}.
\end{align}

\subsection{Complex Kac-Moody current}
By contrast with the undeformed case where the pole of the twist function was second order, here the twist function has only simple poles at $z, \bar z$. Correspondingly, we only expand the Lax matrix to first order there, in other words we evaluate the Lax matrix at the poles of the twist function. Applying Proposition \ref{prop: KM currents} to the case at hand we find that $\L(z, \cdot)$ and $\L(\bar z, \cdot)$ satisfy
\begin{subequations} \label{LL algebra I pre}
\begin{align}
\label{LL algebra Ia pre} \frac{i}{2 \gamma} \big\{ \L_{\1}(z, \theta), \L_{\2}(z, \theta') \big\}_{\sf c} &= - \big[ C_{\1\2}, \L_{\2}(z, \theta) \big] \delta_{\theta \theta'} - C_{\1\2} \delta'_{\theta \theta'},\\
\label{LL algebra Ib pre} \big\{ \L_{\1}(z, \theta), \L_{\2}(\bar z, \theta') \big\}_{\sf c} &= 0,
\end{align}
where we have introduced the real parameter
\begin{equation*}
\gamma = \frac{{\bar z}^2 - z^2}{4 i} = - \Re z \Im z \in \mathbb{R}.
\end{equation*}
Note that $\gamma$ vanishes when the points $z, \bar z$ coalesce, which corresponds to the undeformed limit, and will thus play the role of the deformation parameter in what follows.
By the reality condition \eqref{reality Lax} we have $\tau(\L(z, \theta)) = \L(\bar z, \theta)$ and therefore applying the anti-holomorphic involution $\tau$ to \eqref{LL algebra Ia pre} we also obtain
\begin{equation} \label{LL algebra Ic pre}
- \frac{i}{2 \gamma} \big\{ \L_{\1}(\bar z, \theta), \L_{\2}(\bar z, \theta') \big\}_{\sf c} = - \big[ C_{\1\2}, \L_{\2}(\bar z, \theta) \big] \delta_{\theta \theta'} - C_{\1\2} \delta'_{\theta \theta'}.
\end{equation}
\end{subequations}

Equations \eqref{LL algebra I pre} describe a $\g^{\CC}$-valued Kac-Moody current of level $i/2 \gamma$ which Poisson commutes with its image under $\tau$. Explicitly, we define a field $\Jkm \in \Loop \g^\CC$ as
\begin{equation} \label{KM type I}
\Jkm \coloneqq \frac{i}{2 \gamma} \L(z, \cdot),
\end{equation}
whose complex conjugate is given by $\bar{\Jkm} \coloneqq \tau(\Jkm) = - \frac{i}{2 \gamma} \L(\bar z, \cdot)$. In the complex branch the field $\Jkm \in \Loop \g^\CC$ will play a more fundamental role than the fields $A, \Pi \in \Loop \g$ which are intimately tied with the undeformed theory.
The Poisson brackets between the fields $\Jkm$ and $\bar \Jkm$, equivalent to \eqref{PB A Pi}, are
\begin{subequations} \label{LL algebra I}
\begin{align}
\label{LL algebra Ia} \big\{ \Jkm_{\1}(\theta), \Jkm_{\2}(\theta') \big\}_{\sf c} &= - \big[ C_{\1\2}, \Jkm_{\2}(\theta) \big] \delta_{\theta \theta'} - \mbox{\small $\frac{i}{2 \gamma}$} C_{\1\2} \delta'_{\theta \theta'},\\
\label{LL algebra Ib} \big\{ \Jkm_{\1}(\theta), \bar{\Jkm}_{\2}(\theta') \big\}_{\sf c} &= 0,\\
\label{LL algebra Ic} \big\{ \bar{\Jkm}_{\1}(\theta), \bar{\Jkm}_{\2}(\theta') \big\}_{\sf c} &= - \big[ C_{\1\2}, \bar{\Jkm}_{\2}(\theta) \big] \delta_{\theta \theta'} + \mbox{\small $\frac{i}{2 \gamma}$} C_{\1\2} \delta'_{\theta \theta'}.
\end{align}
\end{subequations}
We extend the action of the automorphism $\sigma : \g \to \g$ to the complexified Lie algebra $\g^\CC$ as $\sigma(X \otimes u) = \sigma(X) \otimes u$ for any $X \in \g$ and $u \in \CC$. Then the decomposition $\g = \g^{(0)} \oplus \g^{(1)}$ extends to $\g^\CC$ in a natural way. We denote by $\Jkm^{(i)}$, $\bar \Jkm^{(i)}$ for $i \in \mathbb{Z}_2$ the components of the field $\Jkm$ and its complex conjugate $\bar \Jkm$ relative to this decomposition.
The Lax matrix \eqref{Lax matrix sym} can then be rewritten in terms of the graded components of the fields $\Jkm$ and $\bar \Jkm$ as
\begin{equation} \label{Lax type I}
\L(\lambda, \cdot) = \ha \big(\lambda^2 - {\bar z}^2\big) \Big( \Jkm^{(0)} + \frac{z}{\lambda} \Jkm^{(1)} \Big) + \ha \big(\lambda^2 - z^2\big) \Big( \bar{\Jkm}^{(0)} + \frac{\bar z}{\lambda} \bar{\Jkm}^{(1)} \Big).
\end{equation}

It is useful to note on passing that the algebra \eqref{LL algebra undef} of the undeformed model can be seen as a degenerate limit of \eqref{LL algebra I pre} when $z \to 1$. Indeed, since $\gamma \to 0$ in this limit, we directly obtain \eqref{LL algebra undef a} from \eqref{LL algebra Ia pre}. Next, taking the difference of \eqref{LL algebra Ia pre} with \eqref{LL algebra Ib pre} we find that
\begin{equation*}
\Big\{ \L_{\1}(z, \theta), \frac{i}{2 \gamma} \big( \L_{\2}(z, \theta) - \L_{\2}(\bar z, \theta) \big) \Big\}_{\sf c} = - \big[ C_{\1\2}, \L_{\2}(z, \theta) \big] \delta_{\sigma \sigma'} - C_{\1\2} \delta'_{\sigma \sigma'}.
\end{equation*}
Therefore in the limit $z \to 1$, noting that $\frac{i}{2 \gamma} \big( \L_{\2}(z, \theta) - \L_{\2}(\bar z, \theta) \big) \to \L'(1, \theta)$, we obtain \eqref{LL algebra undef b}. Finally, taking linear combinations of the brackets \eqref{LL algebra I pre} we also have
\begin{align*}
\Big\{ \frac{i}{2 \gamma} &\big( \L_{\1}(z, \theta) - \L_{\1}(\bar z, \theta) \big), \frac{i}{2 \gamma} \big( \L_{\2}(z, \theta) - \L_{\2}(\bar z, \theta) \big) \Big\}_{\sf c}\\
&\qquad\qquad\qquad\qquad\qquad\qquad = - \Big[ C_{\1\2}, \frac{i}{2 \gamma} \big( \L_{\2}(z, \theta) - \L_{\2}(\bar z, \theta) \big) \Big] \delta_{\sigma \sigma'}.
\end{align*}
In the limit $z \to 1$ this gives rise to \eqref{LL algebra undef c}.

\paragraph{Exchange algebra.} Introduce the field
\begin{equation} \label{Psi def I}
\Psi^{\sf c} \coloneqq \Psi(z, \cdot),
\end{equation}
which takes values in $G^\CC$ since $z$ is complex, and define also $\overline{\Psi^{\sf c}} \coloneqq \tau(\Psi^{\sf c}) = \Psi(\bar z, \cdot)$. The complex Kac-Moody current \eqref{KM type I} and its complex conjugate are expressed in terms of \eqref{Psi def I} as
\begin{equation*}
\Jkm = - \mbox{\small $\frac{i}{2 \gamma}$} \partial_\theta \Psi^{\sf c} (\Psi^{\sf c})^{-1}, \qquad
\bar \Jkm = \mbox{\small $\frac{i}{2 \gamma}$} \partial_\theta \overline{\Psi^{\sf c}} (\overline{\Psi^{\sf c}})^{-1}.
\end{equation*}

To write down the set of Poisson brackets between the fields $\Psi^{\sf c}$ and $\overline{\Psi^{\sf c}}$, let us choose a non-split $R$-matrix $R : \g \to \g$. Consider the corresponding map \eqref{Rpm maps} and its complex conjugate $R^+ = (R + i) : \g_R \to \g^\CC$, whose kernels with respect to the bilinear form on $\g^{\CC}$ we denote $R^{\pm}_{\1\2} = R_{\1\2} \pm i C_{\1\2}$. It is straightforward to show that the Poisson brackets \eqref{LL algebra I} can be deduced from the following (multivalued) Poisson brackets
\begin{subequations} \label{Psi bracket}
\begin{align}
\label{Psi bracket a} \{ \Psi^{\sf c}_{\1}(\theta), \Psi^{\sf c}_{\2}(\theta') \}_{\sf c} &= - \gamma \Psi^{\sf c}_{\1}(\theta) \Psi^{\sf c}_{\2}(\theta') \big( R^+_{\1\2} H_{\theta \theta'} + R^-_{\1\2} H_{\theta' \theta} \big),\\
\label{Psi bracket b} \{ \Psi^{\sf c}_{\1}(\theta), \overline{\Psi^{\sf c}_{\2}}(\theta') \}_{\sf c} &= 0,
\end{align}
where $H_{\theta \theta'}$ is the stair-step function defined by $H_{\theta \theta'} = n$ if $2 \pi n < \theta - \theta' < 2 \pi (n+1)$ and which has the property $\partial_\theta H_{\theta \theta'} = \delta_{\theta \theta'}$. Note that the presence of the non-split $R$-matrix in these brackets is essential to ensure that they are skew-symmetric and satisfy the Jacobi identity. Indeed, the skew-symmetry of \eqref{Psi bracket a} follows at once from the property $R^+_{\1\2} = - R^-_{\2\1}$, which holds by skew-symmetry of $R$. Moreover, one can show that the Jacobi identity for \eqref{Psi bracket} is a direct consequence of the fact that $R$ satisfies mCYBE$(i)$ in tensor form \eqref{mCYBE tensor}. Applying the anti-linear involution $\tau$ to \eqref{Psi bracket a} we also obtain
\begin{equation} \label{Psi bracket abar}
\{ \overline{\Psi^{\sf c}_{\1}}(\theta), \overline{\Psi^{\sf c}_{\2}}(\theta') \}_{\sf c} = - \gamma \overline{\Psi^{\sf c}_{\1}}(\theta) \overline{\Psi^{\sf c}_{\2}}(\theta') \big( R^-_{\1\2} H_{\theta \theta'} + R^+_{\1\2} H_{\theta' \theta} \big).
\end{equation}
\end{subequations}

\subsection{`Yang-Baxter' lift to the cotangent bundle $T^\ast \Loop G$}

In order to define a field $g$ taking values in $G$, consider the non-split $R$-matrix used in writing down the Poisson brackets \eqref{Psi bracket}. In this article we will suppose for simplicity that $\Psi^{\sf c}$ takes values in the main cell $G G_R \subset G^{\CC}$ of the cell decomposition \eqref{GC cell decomp}. We can then decompose $\Psi^{\sf c}(\theta)$ as
\begin{equation} \label{Psi+ decomp}
\Psi^{\sf c}(\theta) = g(\theta) x(\theta)
\end{equation}
for some $g(\theta) \in G$ and $x(\theta) \in G_R$. A similar procedure for extracting the field $g$ of the Yang-Baxter $\sigma$-model on a compact Lie group $G$ from the Iwasawa decomposition of the extended solution at a special value of the spectral parameter (which was identified in \cite{Delduc:2013fga} to be one of the poles of the twist function) has been previously considered in \cite{Klimcik:2008eq}. Applying the anti-linear automorphism $\tau$ to \eqref{Psi+ decomp} we also obtain
\begin{equation} \label{Psi- decomp}
\overline{\Psi^{\sf c}}(\theta) = g(\theta) \bar{x}(\theta)
\end{equation}
where $\bar{x} \coloneqq \tau(x)$ and using $\tau(g) = g$.

In the example of the symmetric space $\sigma$-model we are considering, the parameter $\gamma$ which first appeared in \eqref{LL algebra Ia pre} is real. However, more generally this is obtained from the residue of the twist function at one of its poles, and may therefore be complex. Since $\gamma$ plays the role of the deformation parameter, in this case there are two real deformation parameters, namely $\Re \gamma$ and $\Im \gamma$. We will give an example of this in section \ref{sec: YB+WZW} below. In this more general setting, the Poisson bracket \eqref{Psi bracket abar} should be replaced by
\begin{equation} \label{Psi bracket abar 2}
\{ \overline{\Psi^{\sf c}_{\1}}(\theta), \overline{\Psi^{\sf c}_{\2}}(\theta') \}_{\sf c} = - \bar \gamma \overline{\Psi^{\sf c}_{\1}}(\theta) \overline{\Psi^{\sf c}_{\2}}(\theta') \big( R^-_{\1\2} H_{\theta \theta'} + R^+_{\1\2} H_{\theta' \theta} \big).
\end{equation}
In the following proposition we thus consider the case $\gamma \in \CC$ since it will be applicable more generally. Given a field $y \in \Loop \g^\CC$, we define its imaginary part as 
\begin{equation*}
\Im(y) \coloneqq \frac{1}{2i} \big( y - \tau y \big) \in \Loop \g.
\end{equation*}

\begin{proposition} \label{prop: YB lift I}
Let $g$ and $x$ be defined in terms of $\Psi^{\sf c}$ through the factorisation \eqref{Psi+ decomp}. Introducing $\J \coloneqq - \Im \big( \gamma^{-1} \partial_\theta x x^{-1} \big) \in \Loop \g$, we have the following Poisson brackets
\begin{subequations} \label{PB g X I}
\begin{align}
\label{gg PB I}
\{ g_{\1}(\theta), g_{\2}(\theta') \}_{\sf c} &= 0,\\
\label{Xg PB I}
\{ \J_{\1}(\theta), g_{\2}(\theta') \}_{\sf c} &= g_{\2}(\theta) C_{\1\2} \delta_{\theta \theta'},\\
\label{XX PB I}
\{ \J_{\1}(\theta), \J_{\2}(\theta') \}_{\sf c} &= - [C_{\1\2}, \J_{\2}(\theta)] \delta_{\theta \theta'} - \Im (\gamma^{-1}) C_{\1\2} \delta'_{\theta \theta'}.
\end{align}
If $\gamma \not \in \mathbb{R}$ then we assume that the $R$-matrix satisfies $R^2 = - \textup{id}$.
\end{subequations}
\begin{proof}
By a direct calculation we find that the Poisson brackets \eqref{Psi bracket} can be deduced from the factorisation \eqref{Psi+ decomp} and the following Poisson brackets between the group valued fields $g$, $x$ and $\bar x$,
\begin{subequations} \label{PB x g} 
\begin{align}
\label{xg PB}
\{ x_{\1}(\theta), g_{\2}(\theta') \}_{\sf c} &= \gamma x_{\1}(\theta) g_{\2}(\theta') (R^-_{\1\2})^{x(\theta')} H_{\theta \theta'},\\
\label{xx PB}
\{ x_{\1}(\theta), x_{\2}(\theta') \}_{\sf c} 
&= - \gamma x_{\1}(\theta) \big( (R^-_{\1\2})^{x(\theta')} - (R^-_{\2\1})^{x(\theta')^{-1}} \big) x_{\2}(\theta') H_{\theta \theta'} \notag\\
&\qquad - \gamma x_{\2}(\theta') \big( (R^-_{\1\2})^{x(\theta)^{-1}} - (R^-_{\2\1})^{x(\theta)} \big) x_{\1}(\theta) H_{\theta' \theta},\\
\label{xbx PB}
\{ x_{\1}(\theta), \bar{x}_{\2}(\theta') \}_{\sf c} 
&= - \gamma x_{\1}(\theta) (R^-_{\1\2})^{x(\theta')} \bar{x}_{\2}(\theta') H_{\theta \theta'} + \bar \gamma \bar{x}_{\2}(\theta') (R^+_{\2\1})^{\bar{x}(\theta)} x_{\1}(\theta) H_{\theta' \theta},
\end{align}
\end{subequations}
along with their complex conjugates and \eqref{gg PB I}. Here we have introduced the following shorthand notation
\begin{equation} \label{adxR}
A_{\1\2}^{y(\theta)} \coloneqq y_{\1}(\theta)^{-1} A_{\1\2} y_{\1}(\theta)
\end{equation}
for any $A \in \g^\CC \otimes \g^\CC$ and field $y$ valued in $G^\CC$. It can be checked that \eqref{PB x g} together with \eqref{gg PB I} satisfy the Jacobi identity by virtue once again of \eqref{mCYBE tensor}.

Next, differentiating \eqref{PB x g} with respect to both $\theta$ and $\theta'$ we find that
\begin{align} \label{R2id eq I}
&\{ \partial_\theta x_{\1}(\theta) x_{\1}(\theta)^{-1}, g_{\2}(\theta') \}_{\sf c} = \gamma g_\2(\theta) R^-_{\1\2} \delta_{\theta \theta'},\\
&\{ \partial_\theta x_{\1}(\theta) x_{\1}(\theta)^{-1}, \partial_{\theta'} x_{\2}(\theta') x_{\2}(\theta')^{-1} \}_{\sf c} \notag\\
&\qquad\qquad\qquad = - \gamma \big[ R_{\1\2}, \partial_\theta x_{\1}(\theta) x_{\1}(\theta)^{-1} + \partial_\theta x_{\2}(\theta) x_{\2}(\theta)^{-1} \big] \delta_{\theta \theta'}, \notag\\
&\{ \partial_\theta x_{\1}(\theta) x_{\1}(\theta)^{-1}, \partial_{\theta'} \bar x_{\2}(\theta') \bar x_{\2}(\theta')^{-1} \}_{\sf c} \notag\\
&\qquad\qquad\qquad = - \bar \gamma \big[ R^-_{\1\2}, \partial_\theta x_{\1}(\theta) x_{\1}(\theta)^{-1} \big] \delta_{\theta \theta'} - \gamma \big[ R^-_{\1\2}, \partial_\theta \bar x_{\2}(\theta) \bar x_{\2}(\theta)^{-1} \big] \delta_{\theta \theta'} \notag\\
&\qquad\qquad\qquad\qquad\qquad\qquad\qquad\qquad\qquad\qquad\qquad + \big(\gamma^{-1} - {\bar \gamma}^{-1} \big) R^-_{\1\2} \delta'_{\theta \theta'}. \notag
\end{align}
Finally, it remains to note that by taking suitable linear combinations of these Poisson brackets and using the identity $R^+ - R^- = 2 i$, the field
\begin{equation*}
\J = - \Im \big( \gamma^{-1} \partial_\theta x x^{-1} \big) = \frac{1}{2 i} \big( {\bar \gamma}^{-1} \partial_\theta \bar x {\bar x}^{-1} - \gamma^{-1} \partial_\theta x x^{-1} \big)
\end{equation*}
satisfies \eqref{PB g X I}, as required.

The condition on the $R$-matrix when $\gamma \not \in \mathbb{R}$ follows from equation \eqref{R2id eq I} and using the fact that $\partial_{\theta} x x^{-1} \in \Loop \g_R$ to write $\partial_{\theta} x(\theta) x(\theta)^{-1} = R^- Y(\theta)$ for some $Y \in \Loop \g$. Indeed, in terms of this notation, equation \eqref{R2id eq I} implies
\begin{equation*}
\{ R^- Y_{\1}(\theta), g_{\2}(\theta') \}_{\sf c} = \gamma g_\2(\theta) R^-_{\1\2} \delta_{\theta \theta'}, \qquad
\{ R^+ Y_{\1}(\theta), g_{\2}(\theta') \}_{\sf c} = \bar \gamma g_\2(\theta) R^+_{\1\2} \delta_{\theta \theta'}
\end{equation*}
where the second equation follows by taking the complex conjugate of the first. Taking linear combinations of these and using the fact that $R^+ - R^- = 2 i$ we obtain
\begin{align*}
\{ Y_{\1}(\theta), g_{\2}(\theta') \}_{\sf c} &= - \Im \gamma \, g_\2(\theta) R_{\1\2} \delta_{\theta \theta'} + \Re \gamma \, g_\2(\theta) C_{\1\2} \delta_{\theta \theta'}, \\
\{ R Y_{\1}(\theta), g_{\2}(\theta') \}_{\sf c} &= \Re \gamma \, g_\2(\theta) R_{\1\2} \delta_{\theta \theta'} + \Im \gamma \, g_\2(\theta) C_{\1\2} \delta_{\theta \theta'}.
\end{align*}
Finally, acting with the $R$-matrix on the first tensor factor of the first equation above and comparing the result with the second equation we conclude that for consistency of the Poisson brackets \eqref{PB x g} in the case $\gamma \not \in \mathbb{R}$ the $R$-matrix should satisfy $R^2_{\1\2} = - C_{\1\2}$.
\end{proof}
\end{proposition}

In the case at hand where $\gamma \in \mathbb{R}$, since $x$ takes values in $G_R$, we can write
\begin{equation} \label{def X}
\partial_{\theta} x\, x^{-1} = \gamma R^- X
\end{equation}
for some $X \in \Loop \g_R \coloneqq C^\infty(S^1, \g_R) \simeq \Loop \g$.
Similarly, applying the anti-linear involution $\tau$ to \eqref{def X} we obtain also
\begin{equation} \label{def X bar}
\partial_{\theta} \bar{x}\, \bar{x}^{-1} = \gamma R^+ X.
\end{equation}
It follows that in this case $\J = X$ and the coefficient of the $\delta'$-term in \eqref{XX PB I} vanishes so that \eqref{PB g X I} become
\begin{align*}
\{ g_{\1}(\theta), g_{\2}(\theta') \}_{\sf c} &= 0,\\
\{ X_{\1}(\theta), g_{\2}(\theta') \}_{\sf c} &= g_{\2}(\theta) C_{\1\2} \delta_{\theta \theta'},\\
\{ X_{\1}(\theta), X_{\2}(\theta') \}_{\sf c} &= - [C_{\1\2}, X_{\2}(\theta)] \delta_{\theta \theta'}.
\end{align*}
which are nothing but the canonical Poisson brackets \eqref{PB g X} on $T^\ast \Loop G$.

\subsection{Hamiltonian}

We define the Hamiltonian in the complex branch using the same expression as \eqref{Ham def} but with the twist function replaced by \eqref{twist I}, namely
\begin{equation} \label{Ham def I}
\mathcal H_{\sf c} \coloneqq \qa (\res_{\lambda = 0} - \res_{\lambda = \infty}) \big( \L(\lambda, \cdot) \big| \L(\lambda, \cdot) \big) \varphi_{\sf c}(\lambda) d\lambda.
\end{equation}
Using the explicit form of the Lax matrix \eqref{Lax type I} in terms of the graded components of the complex Kac-Moody current $J$, we find
\begin{align*}
\mathcal H_{\sf c} &= \qa \big( z \Jkm^{(1)} + \bar z \bar \Jkm^{(1)} \big| z \Jkm^{(1)} + \bar z \bar \Jkm^{(1)} \big) + \qa \big( \bar z \Jkm^{(1)} + z \bar \Jkm^{(1)} \big| \bar z \Jkm^{(1)} + z \bar \Jkm^{(1)} \big)\\
&\qquad\qquad\qquad\qquad\qquad\qquad + \qa (z^2 - \bar z^2) \big( \Jkm^{(0)} - \bar \Jkm^{(0)} \big| \Jkm^{(0)} + \bar \Jkm^{(0)} \big).
\end{align*}
In order to express $J$ in terms of the fields $g$ and $X$, we use its definition \eqref{KM type I}, the relation between the Lax matrix and the extended solution \eqref{ext sol} and the factorisation \eqref{Psi+ decomp} of the extended solution at the pole of the twist function to write
\begin{equation} \label{J from g X I}
J = - \frac{i}{2 \gamma} \partial_\theta \Psi^{\sf c} (\Psi^{\sf c})^{-1} = - \frac{i}{2 \gamma} \partial_\theta g g^{-1} - \frac{i}{2} g (R^- X) g^{-1}.
\end{equation}
Substituting this into \eqref{Ham def I} yields the Hamiltonian in terms of $g$ and $X$.

Consider the special case where $z = e^{i \vartheta}$ for some $\vartheta \in \mathbb{R}$. The undeformed limit then corresponds to $\vartheta \to 0$. Writing everything in terms of the original fields $A, \Pi \in \Loop \g$ we find that the Hamiltonian \eqref{Ham def I} can be expressed as
\begin{equation*}
\mathcal H_{\sf c} = \ha \big( A^{(1)} \big| A^{(1)} \big) + \ha \big( \Pi^{(1)} \big| \Pi^{(1)} \big) + \big( A^{(0)} \big| \Pi^{(0)} \big) - \epsilon^2 \big( \Pi^{(0)} \big| \Pi^{(0)} \big),
\end{equation*}
where $\epsilon \coloneqq \sin \vartheta$. Up to a change of sign in $\Pi$ which is due to the different choice of conventions in the expression for the Lax matrix \eqref{Lax matrix sym} compared to \cite{Delduc:2013fga}, this agrees with the Hamiltonian of the deformed symmetric space $\sigma$-model constructed in \cite{Delduc:2013fga}. Moreover, the expressions for $A$ and $\Pi$ in terms of the fields $g$ and $X$ given there are obtained from \eqref{J from g X I} using the expression \eqref{Lax matrix sym} for the Lax matrix. We find
\begin{align*}
A^{(0)} &= P_0 \bigg( - \partial_\theta g g^{-1} + \frac{\eta}{1 + \eta^2} g \big( (R - \eta) X \big) g^{-1} \bigg), \\
A^{(1)} &= \sqrt{1+\eta^2} P_1 \bigg( - \partial_\theta g g^{-1} + \frac{\eta}{1 + \eta^2} g (RX) g^{-1} \bigg),\\
\Pi^{(0)} &= - P_0 \big( g X g^{-1} \big),\\
\Pi^{(1)} &= - \frac{1}{\sqrt{1 + \eta^2}} P_1 \big( g X g^{-1} \big),
\end{align*}
where $\eta \coloneqq \tan \vartheta$. These expressions also agree with \cite{Delduc:2013fga} up to differences in conventions.

\subsection{Two-parameter deformation of the principal chiral model} \label{sec: YB+WZW}

We end this section by giving an example of a two-parameter deformation. Specifically, we shall construct the deformation of the principal chiral model described in \cite{Delduc:2014uaa} in the Hamiltonian setting.

We start from the Lax matrix of the principal chiral model which can be written in terms of the fields $j_0 = - g X g^{-1} \in \Loop \g$ and $j_1 = - \partial_\theta g g^{-1} \in \Loop \g$, where $g \in \Loop G$ and $X \in \Loop \g$ parameterise the cotangent bundle $T^\ast \Loop G$ with the canonical Poisson bracket \eqref{PB g X}, as
\begin{equation*}
\L(\lambda, \cdot) = \frac{1}{1 - \lambda^2} (j_1 + \lambda j_0).
\end{equation*}
The Poisson bracket of the Lax matrix with itself is of the general form \eqref{r-s algebra} where the $\R$-matrix and twist function are given by
\begin{equation*}
\R_{\1\2}(\lambda, \mu) = \frac{C_{\1\2}}{\mu - \lambda} \varphi(\mu)^{-1}, \qquad \varphi(\lambda) = 1 - \frac{1}{\lambda^2}.
\end{equation*}
The twist function has a double pole at the origin, which we deform to two simple poles at $k \pm i A$ in the complex branch. The corresponding deformed twist function reads
\begin{equation*}
\varphi_{\sf c}(\lambda) = \frac{\lambda^2 - 1}{A^2 + (\lambda - k)^2}.
\end{equation*}
Applying Proposition \ref{prop: KM currents} we find that
\begin{equation} \label{2pcm J 1}
J \coloneqq \frac{i}{2 \gamma} \L(k + i A, \cdot) = - \frac{1}{2 i A} \big( j_1 + (k + i A) j_0 \big)
\end{equation}
is a Kac-Moody current of level $\frac{i}{2 \gamma}$ and its complex conjugate $\bar J = \tau(J)$ is a Kac-Moody current of level $- \frac{i}{2 \bar \gamma}$ which Poisson commutes with $J$. Here we have
\begin{equation*}
\gamma = \frac{A}{1 - (k + i A)^2}.
\end{equation*}
Factorising the extended solution at the point $k + i A$ as $\Psi^{\sf c}(\theta) = g(\theta) x(\theta)$ we may write \eqref{ext sol} as
\begin{equation} \label{2pcm J 2}
J = - \frac{i}{2 \gamma} \partial_\theta \Psi^{\sf c} (\Psi^{\sf c})^{-1} = - \frac{i}{2 \gamma} \big( \partial_\theta g g^{-1} + g (\partial_\theta x x^{-1}) g^{-1} \big).
\end{equation}
Combining equations \eqref{2pcm J 1} and \eqref{2pcm J 2} together with their complex conjugates and using the definition in Proposition \ref{prop: YB lift I} of $\J$ in terms of $\partial_\theta x x^{-1}$ and its conjugate, we find
\begin{align*}
j_0 &= g \J g^{-1} + 2 k \partial_\theta g g^{-1}, \\
j_1 &= - k g \J g^{-1} - A g (R\J) g^{-1} - (1 + k^2 + A^2) \partial_\theta g g^{-1}.
\end{align*}
The Hamiltonian \eqref{Ham def} is then given in terms of these currents by
\begin{equation*}
\mathcal H = \frac{(1 + k^2 + A^2) ( j_0 + j_1 | j_0 + j_1 ) + 4 k ( j_0 | j_1) }{4 (A^2 + (k-1)^2)(A^2 + (k+1)^2)}.
\end{equation*}
The Poisson brackets on the fields $g \in \Loop G$ and $\J \in \Loop \g$ parameterising the cotangent bundle are obtained from Proposition \ref{prop: YB lift I} and read
\begin{align*}
\{ g_{\1}(\theta), g_{\2}(\theta') \}_{\sf c} &= 0,\\
\{ \J_{\1}(\theta), g_{\2}(\theta') \}_{\sf c} &= g_{\2}(\theta) C_{\1\2} \delta_{\theta \theta'},\\
\{ \J_{\1}(\theta), \J_{\2}(\theta') \}_{\sf c} &= - [C_{\1\2}, \J_{\2}(\theta)] \delta_{\theta \theta'} + 2 k C_{\1\2} \delta'_{\theta \theta'}.
\end{align*}
The above model coincides up to sign conventions with the one defined in \cite{Delduc:2014uaa} provided we choose an $R$-matrix there which satisfies $R^2 = - \text{id}$.

\section{The real branch} \label{sec: type II}

\begin{figure}[h]
\centering
\def\svgwidth{60mm}
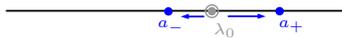
\caption{The real branch.}
\label{fig: poles coset II}
\end{figure}

Another way to deform the twist function $\varphi(\lambda)$ of the symmetric space $\sigma$-model \eqref{coset twist} while preserving the reality conditions is to replace its double pole at $\lambda = 1$ by a pair of simple poles along the real axis, as shown in Figure \ref{fig: poles coset II}.

We therefore define the twist function in the real branch as
\begin{equation} \label{twist II}
\varphi_{\sf r}(\lambda) \coloneqq \frac{4 \lambda}{(\lambda^2 - a_+^2)(\lambda^2 - a_-^2)},
\end{equation}
where $a_{\pm} \in \mathbb{R}$.
As before, the corresponding $\R$-matrix is given by the same expression as in \eqref{scR-matrix} but using the twist function \eqref{twist II}, namely
\begin{equation*}
\R^{\sf r}_{\1\2}(\lambda, \mu) \coloneqq 2 \frac{\lambda C^{(00)}_{\1\2} + \mu C^{(11)}_{\1\2}}{\mu^2 - \lambda^2} \varphi_{\sf r}(\mu)^{-1}.
\end{equation*}
The deformed Poisson bracket is defined by keeping the algebra of Lax matrices of the exact same form as in \eqref{r-s algebra} but using the above $\R$-matrix. That is, we let
\begin{align} \label{r-s algebra II}
\big\{ \L_{\1}(\lambda, \theta), \L_{\2}(\mu, \theta') \big\}_{\sf r} &= \big[\R^{\sf r}_{\1\2}(\lambda, \mu), \L_{\1}(\lambda, \theta)\big] \delta_{\theta \theta'} - \big[\R^{\sf r}_{\2\1}(\mu, \lambda), \L_{\2}(\mu, \theta)\big] \delta_{\theta \theta'} \notag\\
&\qquad\qquad\qquad\qquad\qquad - \big( \R^{\sf r}_{\1\2}(\lambda, \mu) + \R^{\sf r}_{\2\1}(\mu, \lambda) \big) \delta'_{\theta \theta'}.
\end{align}

\subsection{Real commuting Kac-Moody currents}

Applying Proposition \ref{prop: KM currents} to the case at hand we find that the Lax matrix evaluated at the poles $a_\pm$ of the twist function \eqref{twist II} satisfies the following algebra
\begin{subequations} \label{LL algebra II pre}
\begin{align}
\pm \frac{1}{2 \upsilon} \big\{ \L_{\1}(a_{\pm}, \theta), \L_{\2}(a_{\pm}, \theta') \big\}_{\sf r} &= - \big[ C_{\1\2}, \L_{\2}(a_{\pm}, \theta) \big] \delta_{\theta \theta'} - C_{\1\2} \delta'_{\theta \theta'},\\
\big\{ \L_{\1}(a_+, \theta), \L_{\2}(a_-, \theta') \big\}_{\sf r} &= 0,
\end{align}
\end{subequations}
where we have introduced the real parameter
\begin{equation*}
\upsilon \coloneqq \frac{a_+^2 - a_-^2}{4} \in \mathbb{R}.
\end{equation*}
Since this variable vanishes when the points $a_+$ and $a_-$ coalesce, it will play the role of the deformation parameter in what follows. As noted in the complex branch, the Poisson brackets \eqref{LL algebra II pre} degenerate in the limit $a_\pm \to 1$ to the undeformed ones \eqref{LL algebra undef}.

Equations \eqref{LL algebra II pre} describe a pair of $\g$-valued Poisson commuting Kac-Moody currents of level $\pm 1/2 \upsilon$ respectively. Explicitly, let us define fields $\Jkm_\pm \in \Loop \g$ by
\begin{equation} \label{KM type II}
\Jkm_\pm \coloneqq \pm \frac{1}{2 \upsilon} \L(a_\pm, \cdot).
\end{equation}
These replace the pair of fields $A, \Pi \in \Loop \g$ of the undeformed theory in the real branch. Their Poisson brackets, which are equivalent to \eqref{PB A Pi}, follow from \eqref{LL algebra II pre} and read
\begin{subequations} \label{LL algebra II}
\begin{align}
\label{LL algebra IIa} \big\{ \Jkm_{\pm \1}(\theta), \Jkm_{\pm \2}(\theta') \big\}_{\sf r} &= - \big[ C_{\1\2}, \Jkm_{\pm \2}(\theta) \big] \delta_{\theta \theta'} \mp \mbox{\small $\frac{1}{2 \upsilon}$} C_{\1\2} \delta'_{\theta \theta'},\\
\label{LL algebra IIb} \big\{ \Jkm_{+ \1}(\theta), \Jkm_{- \2}(\theta') \big\}_{\sf r} &= 0.
\end{align}
\end{subequations}
We denote by $J_\pm^{(i)}$ for $i \in \mathbb{Z}_2$ the components of $J_\pm$ along the eigenspaces $\g^{(i)} \subset \g$ of the automorphism $\sigma$.
The Lax matrix \eqref{Lax matrix sym} can then be expressed in terms of these as
\begin{equation} \label{Lax type II}
\L(\lambda, \cdot) = \ha \big(\lambda^2 - a_-^2\big) \Big( \Jkm_+^{(0)} + \frac{a_+}{\lambda} \Jkm_+^{(1)} \Big) + \ha \big(\lambda^2 - a_+^2\big) \Big( \Jkm_-^{(0)} + \frac{a_-}{\lambda} \Jkm_-^{(1)} \Big).
\end{equation}

\paragraph{Exchange algebra.}

We define the fields
\begin{equation} \label{Psipm II}
\Psi^{\pm} \coloneqq \Psi(a_{\pm}, \cdot),
\end{equation}
which by virtue of $a_\pm$ being real, both take values in the real Lie group $G$. The analogue of the field \eqref{Psi def I} taking values in the complex double $G^\CC$, in the present case is
\begin{equation} \label{Psi def II}
\Psi^{\sf r} \coloneqq \big( \Psi^+, \Psi^- \big)
\end{equation}
valued in $D$.
The pair of Kac-Moody currents defined in \eqref{KM type II} are expressed in terms of \eqref{Psipm II} as
\begin{equation*}
\Jkm_\pm = \mp \mbox{\small $\frac{1}{2 \upsilon}$} \partial_\theta \Psi^\pm (\Psi^\pm)^{-1}.
\end{equation*}

We are interested in writing down the Poisson brackets of the fields \eqref{Psipm II}. Suppose that we have a split $R$-matrix $R : \g \to \g$ and introduce the corresponding maps \eqref{Rtpm maps} whose kernels with respect to the bilinear form on $\g$ we denote $R^{\pm}_{\1\2} = R_{\1\2} \pm C_{\1\2}$. A direct calculation reveals that the Poisson brackets \eqref{LL algebra II} are a consequence of the following (multivalued) Poisson brackets
\begin{subequations} \label{Psi bracket II}
\begin{align}
\label{Psi bracket II a} \{ \Psi^{\pm}_{\1}(\theta), \Psi^{\pm}_{\2}(\theta') \}_{\sf r} &= \pm \upsilon \Psi^{\pm}_{\1}(\theta) \Psi^{\pm}_{\2}(\theta') \big( R^+_{\1\2} H_{\theta \theta'} + R^-_{\1\2} H_{\theta' \theta} \big) \notag\\
&= \pm \upsilon \Psi^{\pm}_{\1}(\theta) \Psi^{\pm}_{\2}(\theta') \big( R_{\1\2} + C_{\1\2} \epsilon_{\theta \theta'} \big),\\
\label{Psi bracket II b} \{ \Psi^+_{\1}(\theta), \Psi^-_{\2}(\theta') \}_{\sf r} &= 0.
\end{align}
\end{subequations}
As in the complex branch, the skew-symmetry of the $R$-matrix, which can be phrased as $R^+_{\1\2} = - R^-_{\2\1}$, ensures that the Poisson brackets \eqref{Psi bracket II a} are skew-symmetric. Moreover, the fact that $R$ satisfies mCYBE$(1)$, which is written in tensorial notation as \eqref{mCYBE tensor}, ensures that the Poisson brackets \eqref{Psi bracket II} satisfy the Jacobi identity.

\subsection{`Yang-Baxter' lift to the cotangent bundle $T^\ast \Loop G$}

Following the same procedure as used in the complex branch, in order to define a pair of fields $g \in \Loop G$ and $X \in \Loop \g$ satisfying either the Poisson brackets \eqref{PB g X} or \eqref{PB F J}, we shall make use of the same split $R$-matrix that we used to write down the Poisson brackets \eqref{Psi bracket II} of $\Psi^{\sf r} \coloneqq \big( \Psi^+, \Psi^- \big)$. Assuming the latter takes values in the main cell $G^\delta G_R \subset D$ of the cell decomposition \eqref{D cell decomp} for all values of $\theta$, we can factorise it as
\begin{equation*}
\Psi^{\sf r}(\theta) = \big( g(\theta), g(\theta) \big) \big( x_+(\theta), x_-(\theta) \big)
\end{equation*}
for some $g(\theta) \in G$ and $\big( x_+(\theta), x_-(\theta) \big) \in G_R$. In other words, we have
\begin{equation} \label{Psipm decomp}
\Psi^{\pm}(\theta) = g(\theta) x_{\pm}(\theta).
\end{equation}

As we did in the complex branch, in order to prove a more general result we will relax the condition that the levels $\pm \frac{1}{2 \upsilon}$ of the Kac-Moody currents $J_\pm$ are opposite. Indeed, according to Proposition \ref{prop: KM currents} these levels are given by residues of the twist function at its two simple poles so may generically be arbitrary. In what follows we will therefore assume that the Poisson brackets \eqref{Psi bracket II} take the following more general form
\begin{subequations} \label{Psi bracket II 2}
\begin{align}
\label{Psi bracket II 2 a} \{ \Psi^{\pm}_{\1}(\theta), \Psi^{\pm}_{\2}(\theta') \}_{\sf r} &= \pm \upsilon_\pm \Psi^{\pm}_{\1}(\theta) \Psi^{\pm}_{\2}(\theta') \big( R^+_{\1\2} H_{\theta \theta'} + R^-_{\1\2} H_{\theta' \theta} \big)\\
\label{Psi bracket II 2 b} \{ \Psi^+_{\1}(\theta), \Psi^-_{\2}(\theta') \}_{\sf r} &= 0,
\end{align}
\end{subequations}
for some $\upsilon_\pm \in \mathbb{R}$.

\begin{proposition} \label{prop: YB lift II}
Let $g$ and $x_\pm$ be defined in terms of $\Psi^{\sf r}$ by the factorisation \eqref{Psipm decomp}. If we define $\J\coloneqq \ha \big( \upsilon_+^{-1} \partial_\theta x_+ x_+^{-1} - \upsilon_-^{-1} \partial_\theta x_- x_-^{-1} \big) \in \Loop \g$ then we have the following Poisson brackets
\begin{subequations} \label{PB g X II}
\begin{align}
\label{gg PB II}
\{ g_{\1}(\theta), g_{\2}(\theta') \}_{\sf r} &= 0,\\
\label{Xg PB II}
\{ \J_{\1}(\theta), g_{\2}(\theta') \}_{\sf r} &= g_{\2}(\theta) C_{\1\2} \delta_{\theta \theta'},\\
\label{XX PB II}
\{ \J_{\1}(\theta), \J_{\2}(\theta') \}_{\sf r} &= - [C_{\1\2}, \J_{\2}(\theta)] \delta_{\theta \theta'} - \ha (\upsilon_-^{-1} - \upsilon_+^{-1}) \delta'_{\theta \theta'}.
\end{align}
\end{subequations}
If $\upsilon_+ \neq \upsilon_-$ then we assume that the $R$-matrix satisfies $R^2 = \textup{id}$.
\begin{proof}
Using the same notation \eqref{adxR} as in the complex branch, the Poisson brackets \eqref{Psi bracket II} can be shown to follow from the decomposition \eqref{Psipm decomp} together with the following collection of Poisson brackets between the fields $g$ and $x_\pm$,
\begin{subequations} \label{PB x g II}
\begin{align}
\label{xg PB II}
\{ x_{\pm\1}(\theta), g_{\2}(\theta') \}_{\sf r} &= \upsilon_\pm x_{\pm\1}(\theta) g_{\2}(\theta') (R^\pm_{\1\2})^{x_\pm(\theta')} H_{\theta \theta'},\\
\label{xx PB II}
\{ x_{\pm\1}(\theta), x_{\pm\2}(\theta') \}_{\sf r} 
&= \mp \upsilon_\pm x_{\pm\1}(\theta) \big( (R^-_{\2\1})^{x_\pm(\theta')^{-1}} \pm (R^\pm_{\1\2})^{x_\pm(\theta')} \big) x_{\pm\2}(\theta') H_{\theta \theta'} \notag\\
&\qquad \pm \upsilon_\pm x_{\pm\2}(\theta') \big( (R^-_{\1\2})^{x_\pm(\theta)^{-1}} \pm (R^\pm_{\2\1})^{x_\pm(\theta)} \big) x_{\pm\1}(\theta) H_{\theta' \theta},\\
\label{xbx PB II}
\{ x_{+\1}(\theta), x_{-\2}(\theta') \}_{\sf r} 
&= - \upsilon_+ x_{+\1}(\theta) (R^+_{\1\2})^{x_+(\theta')} x_{-\2}(\theta') H_{\theta \theta'} \notag\\
&\qquad\qquad\qquad + \upsilon_- x_{-\2}(\theta') (R^-_{\2\1})^{x_-(\theta)} x_{+\1}(\theta) H_{\theta' \theta},
\end{align}
\end{subequations}
together with \eqref{gg PB II}. These satisfy the Jacobi identity by virtue once again of \eqref{mCYBE tensor}.

Next, differentiating \eqref{PB x g II} with respect to both $\theta$ and $\theta'$ we obtain
\begin{align} \label{R2id II}
&\{ \partial_\theta x_{\pm \1}(\theta) x_{\pm \1}(\theta)^{-1}, g_{\2}(\theta') \}_{\sf r} = \upsilon_\pm g_{\2}(\theta) R^{\pm}_{\1\2} \delta_{\theta \theta'},\\
&\{ \partial_\theta x_{\pm \1}(\theta) x_{\pm \1}(\theta)^{-1}, \partial_{\theta'} x_{\pm \2}(\theta') x_{\pm \2}(\theta')^{-1} \}_{\sf r} \notag\\
&\qquad\qquad = - \upsilon_\pm \big[ R_{\1\2}, \partial_\theta x_{\pm \1}(\theta) x_{\pm \1}(\theta)^{-1} + \partial_\theta x_{\pm \2}(\theta) x_{\pm \2}(\theta)^{-1} \big] \delta_{\theta \theta'}, \notag\\
&\{ \partial_\theta x_{+\1}(\theta) x_{+\1}(\theta)^{-1}, \partial_{\theta'} x_{-\2}(\theta') x_{-\2}(\theta')^{-1} \}_{\sf r} \notag\\
&\qquad\qquad = - \upsilon_- \big[ R^+_{\1\2}, \partial_\theta x_{+\1}(\theta) x_{+\1}(\theta)^{-1} \big] \delta_{\theta \theta'} - \upsilon_+ \big[ R^+_{\1\2}, \partial_\theta x_{- \2}(\theta) x_{- \2}(\theta)^{-1} \big] \delta_{\theta \theta'} \notag\\
&\qquad\qquad\qquad\qquad\qquad\qquad\qquad\qquad\qquad\qquad\qquad\qquad + (v_+ - v_-) R^+_{\1\2} \delta'_{\theta \theta'}. \notag
\end{align}
Finally, taking linear combinations of the above equations to form the expression
\begin{equation*}
\J = \ha \big( \upsilon_+^{-1} \partial_\theta x_+ x_+^{-1} - \upsilon_-^{-1} \partial_\theta x_- x_-^{-1} \big)
\end{equation*}
and using the identity $R^+ - R^- = 2$ we obtain \eqref{PB g X II}.

As in the proof of Proposition \ref{prop: YB lift I}, the condition on the $R$-matrix in the case when $\upsilon_+ \neq \upsilon_-$ comes again from considering the Poisson bracket \eqref{R2id II} and using the fact that we can write $\partial_{\theta} x_\pm(\theta) x_\pm(\theta)^{-1} = R^\pm Y(\theta)$ for some field $Y \in \Loop \g$. Specifically, equation \eqref{R2id II} implies that
\begin{align*}
\{ Y_{\1}(\theta), g_{\2}(\theta') \} &= \frac{\upsilon_+ - \upsilon_-}{2} g_{\2}(\theta) R_{\1\2} \delta_{\theta \theta'} + \frac{\upsilon_+ + \upsilon_-}{2} g_{\2}(\theta) C_{\1\2} \delta_{\theta \theta'},\\
\{ R Y_{\1}(\theta), g_{\2}(\theta') \} &= \frac{\upsilon_+ + \upsilon_-}{2} g_{\2}(\theta) R_{\1\2} \delta_{\theta \theta'} + \frac{\upsilon_+ - \upsilon_-}{2} g_{\2}(\theta) C_{\1\2} \delta_{\theta \theta'}
\end{align*}
Applying $R$ to the first tensor factor of the equation and comparing this with the second equation we conclude that $R$ should satisfy $R^2_{\1\2} = C_{\1\2}$ for consistency when $\upsilon_+ \neq \upsilon_-$.
\end{proof}
\end{proposition}

Consider the special case at hand where $\upsilon_+ = \upsilon_- = \upsilon$. By definition, the Lie algebra of $G_R$ is the image $\iota(\g_R)$ of the embedding $\iota : \g_R \hookrightarrow \d$ defined in section \ref{sec: split case}. Therefore the $G$-valued fields $x_\pm$ satisfy
\begin{equation} \label{def X II}
\partial_{\theta} x_\pm x_\pm^{-1} = \upsilon R^\pm X
\end{equation}
for some $X \in \g_R \simeq \g$. It follows in this case that the field $\J$ of Proposition \ref{prop: YB lift II} is given simply by $\J = X$ and moreover the coefficient of the $\delta'$-term in \eqref{XX PB II} vanishes, so that \eqref{PB g X II} reduce simply to the canonical Poisson brackets \eqref{PB g X} on $T^\ast \Loop G$.

\subsection{`Gauged WZW' lift to the cotangent bundle $T^\ast \Loop G$} \label{sec: type II T-dual}

We note that the Poisson brackets \eqref{Psi bracket II} are identical to the exchange algebras satisfied by chiral WZW fields \cite{BDF90,AS90,F90,G91, Falceto:1992bf}. Indeed, if we identify the left and right chiral WZW fields as $\F_L \coloneqq \Psi^-$, $\F_R \coloneqq \Psi^+$ then the corresponding left and right WZW currents are given explicitly by
\begin{subequations} \label{WZW currents}
\begin{align}
\J_L &\coloneqq \kappa \partial_\theta \F_L\, \F_L^{-1} = \kappa \partial_\theta \Psi^- (\Psi^-)^{-1} = \kappa \L(\mu_-, \cdot),\\
\J_R &\coloneqq - \kappa \partial_\theta \F_R\, \F_R^{-1} = - \kappa \partial_\theta \Psi^+ (\Psi^+)^{-1} = - \kappa \L(\mu_+, \cdot),
\end{align}
\end{subequations}
where we have introduced the parameter $\kappa \coloneqq \frac{1}{2 \upsilon}$. Defining the WZW field as
\begin{equation} \label{WZW field def}
\F \coloneqq \F_L \, \F_R^{-1} = \Psi^- (\Psi^+)^{-1},
\end{equation}
we find that the left and right currents \eqref{WZW currents} are related by
\begin{equation} \label{WZW LR rel}
\J_L = \kappa \partial_\theta \Psi^- (\Psi^-)^{-1} = \kappa \partial_\theta (\F \Psi^+) (\F \Psi^+)^{-1} = \kappa \partial_\theta \F \F^{-1} - \F \J_R \F^{-1}.
\end{equation}
Moreover, the Poisson brackets of the fields $f$, $\J_R$ and $\J_L$ follow from \eqref{Psi bracket II} and read
\begin{subequations} \label{WZW PB}
\begin{align}
\label{FF WZW PB}
\{ \F_{\1}(\theta), \F_{\2}(\theta') \}_{\sf r} &= 0,\\
\label{JRF WZW PB}
\{ \J_{R \1}(\theta), \F_{\2}(\theta') \}_{\sf r} &= \F_{\2}(\theta) C_{\1\2} \delta_{\theta \theta'},\\
\label{JLF WZW PB}
\{ \J_{L \1}(\theta), \F_{\2}(\theta') \}_{\sf r} &= - C_{\1\2} \F_{\2}(\theta) \delta_{\theta \theta'},\\
\label{JRJR WZW PB}
\{ \J_{R \1}(\theta), \J_{R \2}(\theta') \}_{\sf r} &= - [C_{\1\2}, \J_{R \2}(\theta)] \delta_{\theta \theta'} - \kappa \, C_{\1\2} \delta'_{\theta \theta'},\\
\label{JLJL WZW PB}
\{ \J_{L \1}(\theta), \J_{L \2}(\theta') \}_{\sf r} &= - [C_{\1\2}, \J_{L \2}(\theta)] \delta_{\theta \theta'} + \kappa \, C_{\1\2} \delta'_{\theta \theta'},\\
\label{JLJR WZW PB}
\{ \J_{L \1}(\theta), \J_{R \2}(\theta') \}_{\sf r} &= 0.
\end{align}
\end{subequations}
In particular, the pair of fields $f \in \Loop G$ and $\J_R \in \Loop \g$ with Poisson brackets \eqref{FF WZW PB}, \eqref{JRF WZW PB} and \eqref{JRJR WZW PB} describe a lift of the real commuting Kac-Moody current algebras \eqref{LL algebra II} to the cotangent bundle $T^{\ast} \Loop G$ equipped with the modified WZW-type Poisson bracket $\{ \cdot, \cdot \}_\kappa$ as defined in \eqref{PB F J}.

\paragraph{Poisson-Lie $T$-duality.} The deformed model defined by this particular lift to the cotangent bundle $T^\ast \Loop G$ is nothing but the Poisson-Lie $T$-dual, in the sense of \cite{Klimcik:1995ux, Klimcik:1995dy}, of the model defined by the lift to $T^\ast \Loop G$ in Proposition \ref{prop: YB lift II}.

Indeed, suppose that instead of factorising the fields $\Psi^\pm$ as in \eqref{Psipm decomp} according to the decomposition \eqref{D cell decomp}, we use the decomposition of the real double $D$ in the reverse order, namely \eqref{D cell decomp reverse}. Then, assuming again that $\Psi^\pm$ actually takes values in the main cell $G_R G^\delta$, we may write
\begin{equation} \label{Psi+- decomp 2 II}
\Psi^+(\theta) = y_+(\theta) \tilde{g}(\theta), \qquad
\Psi^-(\theta) = y_-(\theta) \tilde{g}(\theta)
\end{equation}
for some $G_R$-valued field $(y_+, y_-)$ and $\tilde{g} \in \Loop G$. Define the field
\begin{equation} \label{F def II}
\F(\theta) := y_-(\theta) y_+(\theta)^{-1},
\end{equation}
which satisfies the reality condition $\tau(\F(\theta)) = y_-(\theta) y_+(\theta)^{-1} = \F(\theta)$, so that in fact we have $\F \in \Loop G$.
The two fields $\Psi^+$ and $\Psi^-$ are then related by
\begin{equation} \label{Psi- F Psi+ II}
\Psi^-(\theta) = \F(\theta) \Psi^+(\theta).
\end{equation}
In other words, $f$ is nothing but the WZW field as defined in \eqref{WZW field def}. Furthermore, the value of the Lax matrix at the poles of the twist function are related through a gauge transformation by $f$, namely
\begin{equation*}
\L(\mu_-, \cdot) = \partial_{\theta} \Psi^- (\Psi^-)^{-1} = \partial_{\theta} \F \F^{-1} + \F \L(\mu_+, \cdot) \F^{-1},
\end{equation*}
which on account of the definitions \eqref{WZW currents} is nothing but the relation \eqref{WZW LR rel}.

\subsection{Hamiltonian}

The Hamiltonian in the real branch is defined by the same expression as \eqref{Ham def} but with the twist function replaced by \eqref{twist II}, namely
\begin{equation} \label{Ham def II}
\mathcal H_{\sf r} \coloneqq \qa (\res_{\lambda = 0} - \res_{\lambda = \infty}) \big( \L(\lambda, \cdot) \big| \L(\lambda, \cdot) \big) \varphi_{\sf r}(\lambda) d\lambda.
\end{equation}
Substituting the explicit form of the Lax matrix \eqref{Lax type II} in terms of the graded components of the real commuting Kac-Moody currents $J_\pm$, we obtain
\begin{align*}
\mathcal H_{\sf r} &= \qa \big( a_+ \Jkm_+^{(1)} + a_- \Jkm_-^{(1)} \big| a_+ \Jkm_+^{(1)} + a_- \Jkm_-^{(1)} \big) + \qa \big( a_- \Jkm_+^{(1)} + a_+ \Jkm_-^{(1)} \big| a_- \Jkm_+^{(1)} + a_+ \Jkm_-^{(1)} \big)\\
&\qquad\qquad\qquad\qquad\qquad\qquad\qquad + \qa (a_+^2 - a_-^2) \big( \Jkm_+^{(0)} - \Jkm_-^{(0)} \big| \Jkm_+^{(0)} + \Jkm_-^{(0)} \big).
\end{align*}

Consider the case where the poles of the twist function are located at $a_\pm = \zeta^{\pm \frac{1}{2}}$ for some $\zeta \in \mathbb{R}$, the undeformed limit corresponding to $\zeta \to 1$. Then the above Hamiltonian becomes
\begin{align*}
\mathcal H_{\sf r} &= \frac{1 - \zeta^2}{4 \zeta} \bigg( \frac{1 + \zeta^2}{1 - \zeta^2} \big( \Jkm_+^{(1)} + \Jkm_-^{(1)} \big| \Jkm_+^{(1)} + \Jkm_-^{(1)} \big) + \frac{4 \zeta}{1 - \zeta^2} \big( \Jkm_+^{(1)} \big| \Jkm_-^{(1)} \big)\\
&\qquad\qquad\qquad\qquad\qquad\qquad\qquad\qquad - \big( \Jkm_+^{(0)} - \Jkm_-^{(0)} \big| \Jkm_+^{(0)} +  \Jkm_-^{(0)} \big) \bigg),
\end{align*}
which is to be compared with the Hamiltonian of the gauged WZW type deformation of the symmetric space $\sigma$-model constructed in \cite{Hollowood:2014rla}.

\section{The CYBE branch} \label{sec: type III}

\begin{figure}[h]
\centering
\def\svgwidth{60mm}
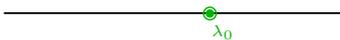
\caption{The CYBE branch.}
\label{fig: poles coset III}
\end{figure}

When the twist function is not deformed, it still has a double pole at $\lambda = 1$ as in the undeformed case, see Figure \ref{fig: poles coset III}, and therefore it is less immediate how to construct a deformation of the $\sigma$-model in this case. However, by using a solution of CYBE one can still perform a canonical transformation by a non-local field to obtain a different model.

Specifically, we want to define a model on the cotangent bundle $T^\ast \Loop G$ parameterised by fields $g \in \Loop G$ and $X \in \Loop G$ so that after a transformation
\begin{equation} \label{can transf}
\widetilde{g} = g k, \qquad \widetilde{X} = k^{-1} X k
\end{equation}
for some field $k$ we recover the undeformed model for the fields $\widetilde{g}$ and $\widetilde{X}$. In other words, we require that the latter are given by the expressions \eqref{g undef} and \eqref{X undef}, respectively. Equivalently, we require that
\begin{align} \label{L1 type III}
\L(1, \cdot) = - \partial_\theta \widetilde{g} \, \widetilde{g}^{-1}, \qquad
\L'(1, \cdot) = - \widetilde{g} \widetilde{X} \widetilde{g}^{-1}.
\end{align}
The motivation for considering a transformation of the type \eqref{can transf} comes from its previous use in the context of $\sigma$-models on Schr\"odinger space-times \cite{Kawaguchi:2013lba} to relate the deformed `exotic symmetry' algebra of these models, identified in \cite{Kawaguchi:2012ug}, to an undeformed (classical) Yangian algebra. Such a transformation was interpreted there as the classical analogue of a Jordanian twist -- see also \cite{Kawaguchi:2014qwa, Matsumoto:2014cja}.

Suppose $R : \g \to \g$ is a solution of CYBE and let $K$ be the subgroup of $G$ with Lie algebra $\k = \text{im} \, R$. We define the non-local field $k$ valued in $K$ to satisfy
\begin{equation}
- \partial_{\theta} k\, k^{-1} = \xi R X,
\end{equation}
for some $\xi \in \mathbb{R}$. Note, in particular, that $k$ need not be periodic. The first two terms \eqref{L1 type III} in the expansion of the Lax matrix at the double pole $\lambda = 1$ of the twist function can then be expressed in terms $g$ and $X$ as
\begin{align*}
\L(1, \cdot) &= - \partial_{\theta} g\, g^{-1} + \xi g (R X) g^{-1}, \\
\L'(1, \cdot) &= - g X g^{-1}.
\end{align*}
Now it is straightforward to check, using the classical Yang-Baxter equation for $R$, that the undeformed Poisson brackets \eqref{LL algebra undef} follow from these relations if the fields $g$ and $X$ satisfy the Poisson brackets \eqref{PB g X}, namely
\begin{subequations} \label{PB g X III}
\begin{align}
\label{PB gg III} \{ g_{\1}(\theta), g_{\2}(\theta') \} &= 0,\\
\label{PB Xg III} \{ X_{\1}(\theta), g_{\2}(\theta') \} &= g_{\2}(\theta) C_{\1\2} \delta_{\theta \theta'},\\
\label{PB XX III} \{ X_{\1}(\theta), X_{\2}(\theta') \} &= - [ C_{\1\2}, X_{\2}(\theta) ] \delta_{\theta \theta'}.
\end{align}
\end{subequations}

\paragraph{Twisted boundary conditions.} Since the field $g \in \Loop G$ defining the deformed model should be periodic, namely $g(2 \pi) = g(0)$, the transformed field $\widetilde{g}$ defined in \eqref{can transf} will in general not be periodic. Indeed, we have
\begin{equation*}
\widetilde{g}(2 \pi) \widetilde{g}(0)^{-1} = g(2 \pi) k(2 \pi) k(0)^{-1} g(0)^{-1} = P \overleftarrow{\exp} \left( - \xi \int_{S^1} g(\theta) \big( R X(\theta) \big) g(\theta)^{-1} d\theta \right).
\end{equation*}
Therefore using the above canonical transformation we can relate the deformed model back to the undeformed one but with twisted boundary conditions \cite{Alday:2005ww}.

\section{Conclusion}

In this article we presented a general procedure for constructing integrable deformations within the Hamiltonian formalism. The starting point for applying the construction is an integrable $\sigma$-model described by a Hamiltonian $\mathcal H$ on the cotangent bundle $T^\ast \Loop G$ and whose twist function $\varphi$, appearing in the Poisson bracket of the Lax matrix $\L$ with itself, has a double pole at some point along the real axis. Typically the Lax matrix only depends on fields parametrising the quotient $G \backslash T^\ast \Loop G$ and as a result the construction consists in two steps. First, the integrable structure is deformed by modifying only the poles of the twist function, resulting in a new twist function $\varphi_\epsilon$. Then, in the second step we construct an integrable $\sigma$-model on $T^\ast \Loop G$ with Hamiltonian $\mathcal H_\epsilon$ defined by the requirement that its integrable structure coincides with the deformed one identified in step one. The second step is achieved with the help of an $R$-matrix. The procedure can be summarised schematically in the diagram below
\begin{equation*}
\xymatrix{
(\mathcal H, T^\ast \Loop G) \ar@{<..>}[d] \ar@{~>}[rrr]^{\text{integrable}}_{\text{deformation}} & & & (\mathcal H_\epsilon, T^\ast \Loop G)\\
(\L, \varphi, G \backslash T^\ast \Loop G) \ar[rrr]^{\text{deform poles of}}_{\text{twist function}} & & & (\L, \varphi_\epsilon, G \backslash T^\ast \Loop G) \ar[u]^{\text{lift with}}_{\text{an } R\text{-matrix}}\\
}
\end{equation*}

The key ingredient used in the second step of the construction is a solution $R \in \text{End}\, \g$ of the modified classical Yang-Baxter equation on the real Lie algebra $\g$. Thus, in order to determine the full list of deformations which can be obtained using this procedure one should turn to the classification of solutions of the mCYBE on real Lie algebras. The classification in the case of complex simple Lie algebras, which is due to Belavin-Drinfel'd \cite{BD}, is very rich. In the case of a real Lie algebra, however, solutions are not as abundant \cite{Cahen}. For instance there are no split $R$-matrices in the case of a compact Lie algebra -- see the Proposition p.12 of \cite{Cahen}. In fact, very few real forms other than the split real form admit solutions of the mCYBE with $c^2 > 0$ \cite[Theorem 3.3]{Cahen}. Unfortunately, this means that the construction of deformations in its present form can only be applied to integrable $\sigma$-models associated to real Lie groups whose Lie algebras admit $R$-matrices of the required type. However, this issue is closely related to the problem of existence of a classical exchange algebra in chiral WZW theory on various real Lie groups which was addressed in \cite{Balog1, Balog2}. As noted there, the problem can be circumvented if one allows the $R$-matrix to depend on the monodromy of the chiral WZW field and require it to satisfy a generalisation of the classical dynamical Yang-Baxter equation \cite{FeherMarshall}. In the present context, one could similarly consider extending the construction to allow for dynamical $R$-matrices which depend explicitly on the monodromy $M \coloneqq \Psi(z, 0)^{-1} \Psi(z, 2 \pi)$ of the extended solution \eqref{ext sol intro} at the poles of the twist function.
Such a generalisation of the present construction will be important in view of describing the complete landscape of possible integrable deformations for a given integrable $\sigma$-model.

\medskip

Even when an $R$-matrix exists on the desired real form, the corresponding Drinfel'd double $\mathcal D$ will not be a bicrossproduct in general. In other words, instead of a simple factorisation $\mathcal D = G G^\ast = G^\ast G$ we will have a cell decomposition as in \eqref{GC cell decomp} or \eqref{D cell decomp}. In this paper we have implicitly assumed that the Drinfel'd double is a bicrossproduct so that the fields parametrising the cotangent bundle $T^\ast \Loop G$ could be extracted by simply factorising the extended solution at the poles of the twist function. Thus an interesting question is whether the construction presented here can be generalised to the case where $\mathcal D$ is not a bicrossproduct but only admits a cell decomposition.

\medskip

Aside from the technicalities mentioned above, there are various interesting directions in which to generalise the construction. In the case of double deformations where $\gamma \not \in \mathbb{R}$ in the complex branch (resp. $\upsilon_+ \neq \upsilon_-$ in the real branch), we were only able to construct deformations using $R$-matrices for which $R^2 = - \text{id}$ (resp. $R^2 = \text{id}$). A natural question is whether the construction of double deformations extends also to other types of $R$-matrices. We note for instance that a slightly weaker condition was used to define a two parameter deformation of the principal chiral model in \cite{Delduc:2014uaa}, namely $R^3 = - R$. It would also be interesting to generalise the formalism so as to enable the construction of bi-Yang-Baxter type models \cite{Klimcik:2014bta}. A further possible direction for generalisation is to consider integrable deformations of integrable $\sigma$-models whose phase space is not given by a cotangent bundle $T^\ast \Loop G$. In this case it will be interesting to identify which object is to play the role of the Drinfel'd double in the present construction.

\subsection*{Acknowledgements}

I would like to thank M. Magro and F. Delduc for useful discussions during initial stages of this project and for comments on the draft.

\providecommand{\href}[2]{#2}\begingroup\raggedright\endgroup

\end{document}